\documentclass[amsmath,amssymb,aps,pre,floatfix,reprint]{revtex4-2}
\usepackage{graphicx}
\usepackage{dcolumn}
\usepackage{bm}
\usepackage{tikz}

\allowdisplaybreaks

\usepackage[subrefformat=parens,labelformat=parens,caption=false]{subfig} 

\usepackage[colorlinks=true]{hyperref}

\begin{document}

\title{Profiling a soft solid layer to passively control the conduit shape in a compliant microchannel during flow}

\author{Pratyaksh Karan}
\affiliation{Department of Mechanical Engineering, Indian Institute of Technology, Kharagpur, West Bengal 721302, India}

\author{Jeevanjyoti Chakraborty}
\thanks{Corresponding author.}
\email{jeevan@mech.iitkgp.ac.in}
\email{jeevanjyoti4@gmail.com}
\affiliation{Department of Mechanical Engineering, Indian Institute of Technology, Kharagpur, West Bengal 721302, India}

\author{Suman Chakraborty}
\affiliation{Department of Mechanical Engineering, Indian Institute of Technology, Kharagpur, West Bengal 721302, India}

\author{Steven T. Wereley}
\affiliation{School of Mechanical Engineering, Purdue University, West Lafayette, Indiana 47907, USA}

\author{Ivan C. Christov}
\thanks{Corresponding author.}
\email{christov@purdue.edu}
\homepage{http://tmnt-lab.org}
\affiliation{School of Mechanical Engineering, Purdue University, West Lafayette, Indiana 47907, USA}

\date{\today}

\begin{abstract}
The shape of a microchannel during flow through it is instrumental to understanding the physics that govern various phenomena ranging from rheological measurements of fluids to separation of particles and cells. Two commonly used approaches for obtaining a desired channel shape (for a given application) are (i) fabricating the microchannel in the requisite shape and (ii) actuating the microchannel walls during flow to obtain the requisite shape. However, these approaches are not always viable. We propose an alternative, passive approach to {\it a priori} tune the elastohydrodynamics in a microsystem, towards achieving a pre-determined (but not pre-fabricated) flow geometry when the microchannel is subjected to flow. That is to say, we use the interaction between a soft solid layer, the viscous flow beneath it and the shaped rigid wall above it, to tune the fluid domain's shape. Specifically, we study a parallel-wall microchannel whose top wall is a slender soft  coating of arbitrary thickness attached to a rigid platform. We derive a nonlinear differential equation for the soft coating's fluid--solid interface, which we use to infer how to achieve specific conduit shapes during flow. Using this theory, we demonstrate the tuning of four categories of microchannel geometries, which establishes, via a proof-of-concept, the viability of our modeling framework. We also explore slip length patterning on the rigid bottom wall of the microchannel, a common technique in microfluidics, as an additional `handle' for microchannel shape control. However, we show that this effect is much weaker in practice.
\end{abstract}

\maketitle

\section{Introduction}\label{sec:intro}

An ubiquitous component of micro-electro-mechanical systems (MEMS) \cite{HT98}, which finds place in applications spanning miniaturized chemical analysis systems \cite{Rossier1999,Gurkan2011} (in micro-total analysis systems, or $\mu$-TAS \cite{RDIAM02,Ahn2004}) to complex fluid rheometry \cite{Degre2006,Pipe2009,DelGiudice2015}, is the microchannel. Having a requisite shape of the microchannel under flow is instrumental for studies on morphology and detection of cells/particles/bubbles \cite{Gurkan2011,Wong2003,Dhong2018}, viscoelasticity of complex fluids \cite{Degre2006,Pipe2009,DelGiudice2015,Zografos2016}, nano- and micro-particle segregation \cite{Chen2019}, amongst other applications \cite{SSA04}. A constriction in a blood vessel often leads to accumulation of plaque on the vessel wall, and therefore obtaining a constricted microchannel shape in an \textit{ex vivo} analysis on a lab-on-a-chip device is of interest \cite{HOLDSWORTH1995,Venugopal2018,Karan2020b}. Narrowing of a microchannel can have a `stretching' effect on cells, vesicles and the like, an aspect of their behavior that is actively being researched \cite{CHUNCHENG2009,Yaginuma2013,Rodrigues2016,Alsmadi2017}. A microchannel's shape has significant bearing on the internal flow's extensional rate, a crucial variable in the characterization of the viscoelasticity of complex fluids \cite{Campo2011,Zografos2016}. Going further, the expansion of microchannels due to the hydrodynamic forces within can be used to control the polydisperisty of emulsions created by co-flow \cite{Pang2014,Zeng2015} and for segregating particles of different sizes via `nano-sieves' \cite{Chen2019}, by simply controlling the inlet pressure to tune the flow geometry in real time.

However, obtaining a desired shape for a microchannel under flow conditions is a non-trivial task due to, in no small part, the potential for flow-induced deformation of a soft wall \cite{Gervais2006}. Two approaches are typically employed in practice: (i) fabricating a close-to-rigid microchannel with a pre-determined shape \cite{Rodd2005,Sun2010,Mulligan2011,Zografos2016}, and (ii) fabricating a soft microchannel that is then actuated by external stimulation during flow to achieve a desired shape for the channel \cite{Ismagilov2001,Chakraborty2012a,Raj2016a,Boyko2018,Boyko2020}. While these approaches are useful and elegant, they are not always viable. For instance, actuating the microchannel at the time of flow can be challenging if the application requires the microchannel to be undergoing motion when it is subject to flow, for example, in a lab-on-a-CD device \cite{Madou2006,Abhimanyu2016}.

A third approach, which has often been overlooked, is: (iii) \textit{a priori} attunement of the elastohydrodynamics (EHD) of a microchannel to induce a desired shape upon achieving steady flow. The prime advantage of this approach, over (i), is that the channel can be fabricated with simplistic geometry (like constant-gap slit geometry or a constant-radius cylindrical geometry), but will assume requisite shape due to EHD when subjected to flow. Likewise, the prime advantage of this approach, over (ii), is that there is no requirement to physically access and handle the microchannel setup when it is in operation, i.e., the operation is hands-off. These advantages make approach (iii) viable and useful alternative to approaches (i) and (ii), in certain situations. Possibly, the reason that a gap in the literature pertaining to this third approach exists is that the discipline of EHD \cite{Gohar2001} in microsystems, of which fluid--structure interactions at low Reynolds number is but one example \cite{DS14}, is in an inchoate stage. Therefore, beyond the aforementioned studies that have focussed on actively actuating microchannels, EHD in microchannels has been studied only to account for its influence \cite{HJLK02,Del2016}, rather than to exploit its presence for a desired outcome. Nevertheless, there is significant interest in such \emph{soft interface} problems, from a fundamental transport phenomena perspective \cite{Stone2017}. Furthermore, recently, `peeling' mechanisms of EHD have been demonstrated to allow for shape control of elastic membranes actuated by fluid flow \cite{Peretz2020,Salem2020}. The passive approach to \textit{a priori} attunement of EHD is particularly relevant within the scope for tunability of common microfluidics materials, such as hydrogels \cite{Huang2011,Kruss2012,Bertassoni2014,Zhao2016,Song2018}. Hydrogels are an emerging class of material for biomimetics and artificial tissue engineering, a common application area of elastohydrodynamics \cite{Huang2011,Bertassoni2014,Kruss2012}. In the last decade, the question of what is the shape of a soft microchannel wall that has been deformed by a steady viscous fluid flow has been explored \cite{Karan2018}: from studies based on scaling correlations \cite{Gervais2006,Raj2016,Raj2017}, to solving the fluid--structure interaction problem in two-dimensional and axisymmetric configurations \cite{Mahadevan2004,Mukherjee2013,Raj2018}, to solution of three-dimensional problems that determine the effect of lateral clamping of the channel walls \cite{Christov2018,Wang2019}, to studies accounting for non-Newtonian fluid rheology \cite{Raj2016,Raj2018,Anand2019}. 

Here, we propose to harness this new understanding provided by the latter fundamental studies to enable \textit{a priori} attuning of EHD in microsystems to recover a desired shape for the flow passage in a microchannel, under steady flow. To this end, we study an infinite parallel plate (two-dimensional) microchannel whose top wall is a soft coating of arbitrary thickness attached to a rigid platform. We propose to achieve a desired axial variation of the microchannel top wall (i.e., the fluid--solid interface) due to  the hydrodynamic forces under imposed flow by controlling the thickness of the soft coating via the bounding rigid platform's shape. 

In addition, we assess slip length patterning on the rigid bottom wall, as is common in microfluidics \cite{Stroock2002,Stroock2003}, as another `handle' to control the system's behavior. Although we will show that slip length patterning has limited use for controlling the microchannel shape under flow, it is a promising approach to modulating the bottom-wall shear rate, which has significant influence on particle migration and physiological processes in flow~\cite{Leighton1987,Phillips1992,Salsac2006,Sebastian2018}. For example, the shear rate is known to have an impact on behavior of biological cells under flow, leading to phenomena like detachment from the wall, chemical release from cells, alterations in platelet function, etc.~\cite{Tang2012shear,Mascari2003,Alsmadi2017}. For instance, it can be desirable to enhance bacterial cell detachment in therapy based on displacement by antibodies \cite{Mascari2003}. On the other hand, in a bio-mimetic studies of cell migration for angiogenesis in tumor micro-environments, it is crucial that the shear rate does not play a role (i.e., that it does not affect the migration of cells), so that their response to chemical gradients can be delineated \cite{Priyadarshani2021}.

To these ends, the remainder of this work is organized as follows. In section \ref{sec:model}, we describe the physical problem that we are studying, including the requisite notation for its mathematical treatment. In section \ref{sec:math}, we formulate the problem mathematically. Specifically, starting from governing equations and boundary conditions (given in appendix \ref{subsec:gdesbcs}), a scaling analysis (section~\ref{subsec:scaling}) leads to the lubrication approximation, and the model is simplified asymptotically to two coupled equations (section \ref{subsec:solution}) --- an ODE for the fluid pressure and a linear algebraic expression for the wall deflection. A set of inverse solutions is formulated in section \ref{subsec:inverse}. We briefly discuss the numerical approach for solving the obtained system of equations in section \ref{subsec:soln}. In section \ref{sec:results}, we present our numerical results on passive control, focussing on four desired canonical shapes under flow. We conclude our study in section \ref{sec:conclusion}, highlighting its salient features and avenues for future work. 

The major outcome of our study, which also demonstrates the application-worthiness of our theoretical framework, is comprised by two complementary components. The first component is a simplified mathematical description of the system behavior in terms of an ordinary differential equation (ODE) for the pressure. The second component is the accompanying detailed characterization of `inverse problems' based on this ODE, which take in the desired microchannel fluid--solid interface shape and bottom wall shear-rate as the input, and return the required solid layer profiling and slip length patterning, respectively, as the output.


\section{Model Setup}\label{sec:model}
\begin{figure}
\centering
\includegraphics[width=0.45\textwidth]{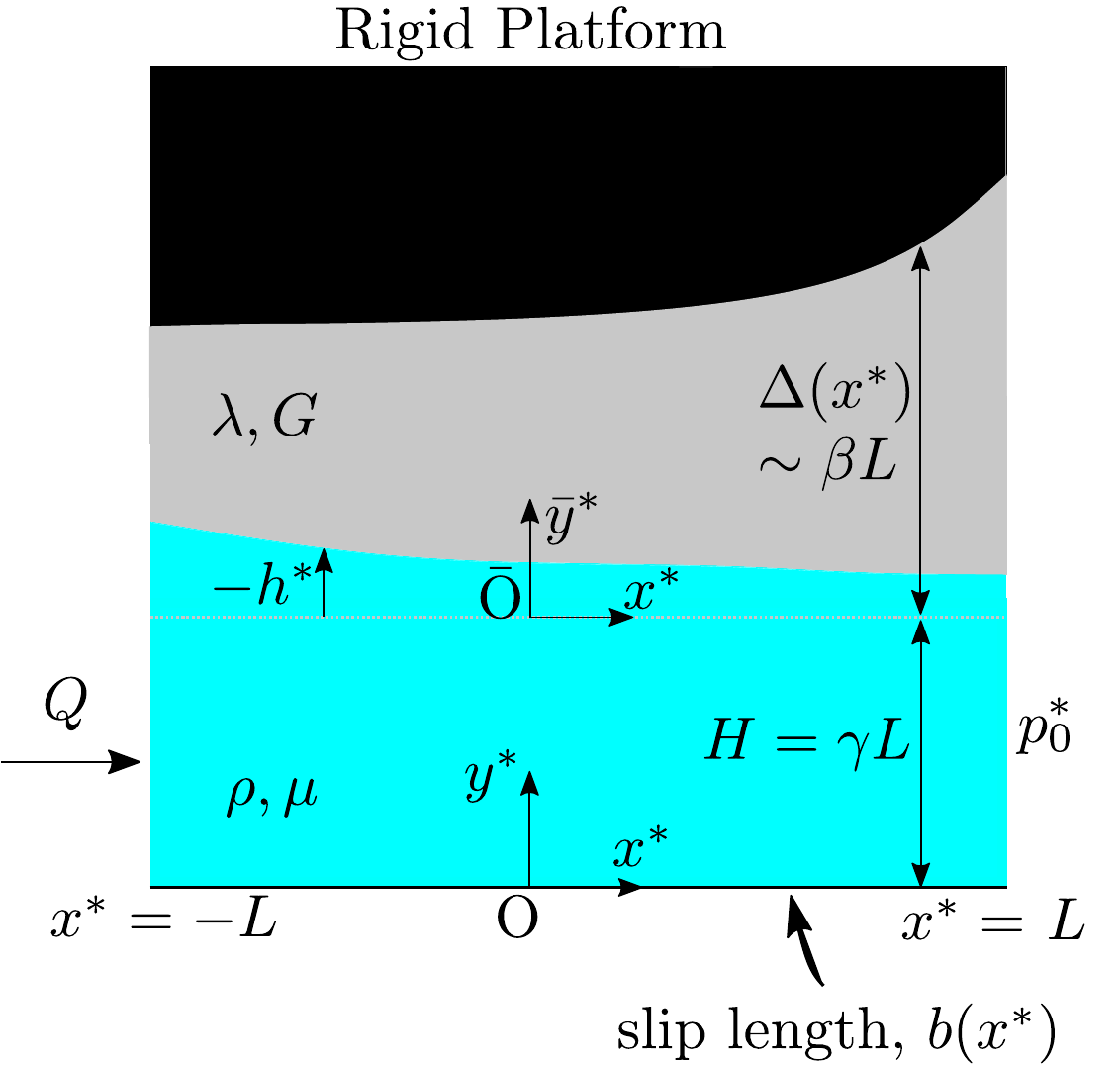}
\caption{(Color online.) Schematic representation of the physical problem setup. The blue region is the flow conduit (fluid domain), the grey region is the elastic layer (solid domain), and the black region is the confining rigid platform. The solid domain has an initial variable thickness along the flow-wise, $x^*$, direction. Due to fluid--structure interaction, each domain's thickness varies with $x^*$ upon achieving steady conditions.}
\label{fig:schematic}
\end{figure}

The physical setup for this study is presented in Figure \ref{fig:schematic}. The `into-the-paper' width of the geometry is considered to be substantially larger than its length in the flow-wise, $x^*$, direction, such that the setup is effectively two-dimensional (2D). We consider the flow of an incompressible Newtonian fluid of density $\rho$ and dynamic viscosity $\mu$ in the initially-rectangular microchannel. The bottom wall of the microchannel has been patterned to generate hydrodynamic slip, which is captured by the linear Navier model \cite{Lauga2003,Lauga2007} with slip length $b(x^*)$, which varies along the flow-wise direction (i.e., along the channel's length). The region between the rigid bottom wall and the deformed elastic interface (which represents the top wall of the microchannel) is referred to as the `fluid domain'. The top wall of the channel constitutes the interface with a layer of linearly elastic solid material with first Lam\'e parameter $\lambda$ and shear modulus $G$. The solid layer is attached to a rigid  platform on top. The interface of the solid layer with the rigid platform varies with $x^*$, i.e., the solid layer has a varying thickness along the flow-wise direction. This variable-thickness layer is referred to as the `solid domain'. 

Two co-ordinate systems, $\mathrm{O}$ and $\mathrm{\bar{O}}$, are employed to avoid ambiguity: one for the fluid domain, $(x^*,y^*)$ attached to the rigid bottom wall, and one for the solid domain, $(x^*,\bar{y}^*)$ attached to the undeformed fluid--solid interface. Note that the vertical axis is $y^*$ in the fluid domain and $\bar{y}^*$ in the solid domain. The horizontal axis is denoted by $x^*$ for both domains. We denote the flow velocity field by $\vec{v}^*(x^*,y^*)$ and the solid displacement field by $\vec{u}^*(x^*,\bar{y}^*)$, with the individual components expressed employing appropriate subscripts. The horizontal extent of both the domains is from $x^*=-L$ at the inlet (on the left) to  $x^*=L$ at the outlet (on the right). The undeformed thickness of the fluid domain is $H$, such that $H/L=\gamma$ is the aspect ratio. The undeformed thickness of the solid layer is $\Delta(x^*)$ such that $\max_{x^*}\Delta(x^*)/ L = \beta$. The volumetric flow rate per unit width is $Q$, and the hydrodynamic pressure at the outlet is $p_0^*$. We restrict our analysis to infinitesimal strains in the solid. We assess the steady-state response of the system, which corresponds to steady flow in the fluid domain and a suitably equilibrated deformation in the solid domain.


\section{Mathematical Formulation}\label{sec:math}

The flow in the fluid domain is governed by the 2D continuity and incompressible Navier Stokes equations, subject to Navier slip and no penetration conditions at the rigid bottom wall, no-slip and no-penetration conditions at the deformed fluid--solid interface, and imposed pressure at the outlet. At steady state, the continuity equation requires that the flow rate is equal across any two axial cross-section and equal to the imposed flow rate at the inlet. The deformation of the solid domain is governed by the equilibrium equations of linear elasticity, subject to a zero-displacement condition at the solid--platform interface and a traction-balance condition at the fluid--solid interface. All the governing equations and requisite boundary conditions and constraints are given in appendix \ref{subsec:gdesbcs}.

\begin{table}
\def\arraystretch{1.25}
\caption{Pertinent dimensionless quantities describing the system and their definitions.}
\label{tab:nondim}
\begin{ruledtabular}
\begin{tabular}{lp{2in}}
\textbf{Parameter}											&	\textbf{Definition}\\
\hline
$\displaystyle \gamma$		   				&	$\displaystyle {H}/{L}$\\[5pt]
$\displaystyle \beta$						&	$\displaystyle {\max_{x^*}\Delta(x^*)}/{L}$\\[5pt]
$\displaystyle \kappa$ 						&	$\displaystyle {x^*_c}/{L}$\\[5pt]
$\displaystyle \phi_0$				 		&	$\displaystyle \frac{\beta\kappa}{\gamma^3}\frac{\mu Q}{(\lambda+2G)L^2}$\rule{0pt}{4ex} \\[5pt]
$\displaystyle \xi(x) = {1}/{\bar{\phi}(x)}$			&	$\displaystyle \frac{\Delta(x^*)}{\max_{x^*}\Delta(x^*)} = \frac{\Delta(x)}{\max_{x}\Delta(x)}$\rule{0pt}{4ex} \\[5pt]
\end{tabular}
\end{ruledtabular}
\end{table}

\subsection{Scaling Analysis}\label{subsec:scaling}

Since the fluid domain is subjected to slip length patterning at the bottom wall, and the solid domain's thickness varies axially as well, the $x^*$-scale will be the smallest of the three available length scales. Since the geometric axial scale of the system is $L$, we take the scales for $x^*$ to be $x^*_c=\kappa L$, where $\kappa$ is to be determined for each physical situation, depending on the particular slip length patterning and solid layer profiling imposed. 

The scale for $y^*$ is $H = \gamma L$, and the scale for  $\bar{y}^*$ is $\beta L$. Importantly, we assume that the solid domain's thickness, over the entire axial length, is substantially smaller that the axial scale of the system, i.e., $\beta \ll \kappa$. The $v_x^*$-scale is given by the mean axial speed at inlet, i.e., ${Q}/({\gamma L})$. Subsequently, the $v_y^*$-scale is found to be ${Q}/({\kappa L})$ by balancing the continuity equation~\eqref{eq:continuity}. From lubrication theory \cite{Leal2007,Stone2017}, we expect that the axial pressure gradient balances the viscous forces in this microflow. Then, scaling the pressure and viscous terms in equation \eqref{eq:xmom} gives us $({\kappa}/{\gamma^3})({\mu Q}/{L^2})$ as the $p^*$-scale. In this scaling, $({\gamma}/{\kappa})({\rho Q}/{\mu})$ plays the role of the (reduced, or lubrication) Reynolds number, and $\gamma/\kappa \ll 1$ is the slenderness parameter for the fluid mechanics problem.
We take the scale for $u_x^*$, $u_y^*$ and $h^*$ to all be $\phi_0 L$, where $\phi_0$ is obtained self-consistently from equation~\eqref{eq:tracbal_y_nd1_reit} below from the traction balance~\eqref{eq:tracbal}. The expressions of the pertinent dimensionless quantities are summarized in table~\ref{tab:nondim}.

In appendix~\ref{subsec:gdesbcs_nd}, we make the governing equations and boundary conditions (from appendix \ref{subsec:gdesbcs}) dimensionless with the characteristic scales discussed above. The dimensionless variables retain the same notation as the dimensional variables, but with the superscript $^*$ dropped.

\subsection{Lubrication Approximation and Asymptotic Reduction}\label{subsec:solution}
As is common in microfluidics \cite{SSA04}, we make the lubrication approximation \cite{Leal2007,Stone2017}. This slenderness assumption works the same way in both the fluid and solid domains. This approximation allows us, as we will see ahead, to obtain a generalized Winkler-like relation between pressure on the solid layer and the layer's deformation. To further justify the approximations made, we give a concise discussion, in appendix \ref{sec:comparison}, on the validity of our proposed modeling framework.

Noting that the fluid--solid interface deflection is primarily in the $y$-direction (or, equivalently, $\bar{y}$-direction), it follows that the $\bar{y}$-component of the traction balance condition (equation \eqref{eq:tracbal_y_nd1}), re-iterated below,
\begin{multline}
\label{eq:tracbal_y_nd1_reit}
\frac{\partial u_{\bar{y}}}{\partial \bar{y}}+ \frac{\beta}{\kappa}\left[\left(\frac{\lambda}{\lambda+2G}\right)\frac{\partial u_x}{\partial x} \right. \\ + \left. \left(\frac{G}{\lambda+2G}\right)\left(\frac{\phi_0}{\kappa}\frac{\partial u_{\bar{y}}}{\partial x}+\frac{\phi_0}{\beta}\frac{\partial u_x}{\partial \bar{y}}\right)\frac{\partial h}{\partial x}\right]  \\
= - \frac{\beta\kappa}{\gamma^3\phi_0}\frac{\mu Q}{(\lambda+2G)L^2} \\ \times \left[p - \frac{\gamma}{\kappa}\left\{\frac{2\gamma}{\kappa}\frac{\partial v_y}{\partial y}+\frac{\phi_0}{\kappa}\left(\frac{\partial v_x}{\partial y}+\frac{\gamma^2}{\kappa^2}\frac{\partial v_y}{\partial x}\right)\right\}\right], \\
\text{at }y = 1-\frac{\phi_0 h(x)}{\gamma},\quad \bar{y} = -\frac{\phi_0 h(x)}{\beta}\approx 0.
\end{multline}
should be balanced asymptotically. In other words, the force from the fluid domain and the force from the solid domain should scale the same way. Hence, we scale the leading-order contribution of the left-hand and right-hand sides of equation \eqref{eq:tracbal_y_nd1_reit} equally, and obtain
\begin{equation}
\phi_0 = \frac{\beta\kappa}{\gamma^3}\frac{\mu Q}{(\lambda+2G)L^2}.
\end{equation}

Now, under the lubrication approximation, we retain only the leading-order terms and obtain the simplified version of equations \eqref{eq:xmom_nd1}, \eqref{eq:ymom_nd1}, \eqref{eq:mecheq_x_nd1}, \eqref{eq:mecheq_y_nd1}, \eqref{eq:tracbal_x_nd1} and~\eqref{eq:tracbal_y_nd1}:  
\begin{align}
\label{eq:xmom_nd2}
0&=-\frac{\partial p}{\partial x}+\frac{\partial^2 v_x}{\partial y^{2}},\\
\label{eq:ymom_nd2}
0&=-\frac{\partial p}{\partial y}.\\
\label{eq:mecheq_x_nd2}
\frac{\partial^2 u_x}{\partial \bar{y}^2}&= 0,\\
\label{eq:mecheq_y_nd2}
\frac{\partial^2 u_{\bar{y}}}{\partial \bar{y}^2} &= 0,\\
\label{eq:tracbal_x_nd2}
\left.\frac{\partial u_x}{\partial \bar{y}}\right|_{\bar{y}=0} &=0,\\
\label{eq:tracbal_y_nd2}
\left.\frac{\partial u_{\bar{y}}}{\partial \bar{y}}\right|_{\bar{y}=0} &= -p.
\end{align}
Equations \eqref{eq:continuity_nd1}, \eqref{eq:NavierSlipbottom_nd1}, \eqref{eq:nopenbottom_nd1}, \eqref{eq:outletp_nd} \eqref{eq:flowconst_nd}, \eqref{eq:noslipnopentop_nd1} and \eqref{eq:nodisptop_nd1} remain unchanged by this approximation.

Now, equation \eqref{eq:xmom_nd2} can be integrated and subjected to the boundary conditions in \eqref{eq:NavierSlipbottom_nd1} and \eqref{eq:noslipnopentop_nd1} to yield: 
\begin{equation}
\label{eq:vx}
v_x(x,y) = \frac{1}{2}\frac{dp}{dx}\left[y^2 -\frac{\left(1-\frac{\phi_0 h(x)}{\gamma}\right)^2\left(y+\frac{b(x)}{\gamma L}\right)}{\left(1-\frac{\phi_0 h(x)}{\gamma}+\frac{b(x)}{\gamma L}\right)}  \right].
\end{equation}
Note that equation \eqref{eq:ymom_nd2} implies that $p$ is no longer explicitly dependent on $y$; thus, $p=p(x)$ only in equation \eqref{eq:vx} as well as the rest of the analysis ahead.

From equation \eqref{eq:vx}, we also obtain expressions for shear rate in the flow:
\begin{equation}
\label{eq:shear}
\frac{\partial v_x}{\partial y} = \frac{1}{2}\frac{dp}{dx}\left[2y-\frac{\left(1-\frac{\phi_0 h(x)}{\gamma}\right)^2}{\left(1-\frac{\phi_0 h(x)}{\gamma}+\frac{b(x)}{\gamma L}\right)}\right].
\end{equation}

Substituting the expression for $v_x$ from equation \eqref{eq:vx} into equation \eqref{eq:flowconst_nd}, we obtain a first-order ordinary differential equation (ODE) for the pressure: 
\begin{multline}
\label{eq:ode_p}
\left(1-\frac{\phi_0 h(x)}{\gamma}\right)^3\left(1-\frac{\phi_0 h(x)}{\gamma}+\frac{4b(x)}{\gamma L}\right)\frac{dp}{dx} \\ + 12\left(1-\frac{\phi_0 h(x)}{\gamma}+\frac{b(x)}{\gamma L}\right) = 0.
\end{multline}

Next, equations \eqref{eq:mecheq_x_nd2} and \eqref{eq:mecheq_y_nd2} are integrated and subjected to the boundary conditions \eqref{eq:nodisptop_nd1}, \eqref{eq:tracbal_x_nd2} and \eqref{eq:tracbal_y_nd2} to yield:
\begin{align}
\label{eq:ux}
u_x(x,\bar{y}) &= 0,\\
\label{eq:uy}
u_{\bar{y}}(x,\bar{y}) &= \big(\xi(x)-\bar{y}\big)p(x).
\end{align}
These equations provide us with the relationship between the fluid--solid interface deflection $h(x)$, which is equal to $-u_{\bar{y}}$ evaluated at $\bar{y}=0$, and the hydrodynamic pressure $p(x)$ as, 
\begin{equation}
\label{eq:h}
h(x) = -\xi(x) p(x) \quad\implies\quad p(x) = -\bar{\phi}(x)h(x),
\end{equation}
where we have introduced the notation $\bar{\phi}(x) \equiv 1/\xi(x)$. Generalizing previous results \cite{Mahadevan2004,Mukherjee2013} to the case of axially-varying confinement, equation~\eqref{eq:h} is essentially with a Winkler-like pressure--deformation relation \cite{Dillard2018} with an axially-varying dimensionless stiffness $\xi(x)$. 

Now, substituting $h$ from equation \eqref{eq:h} into equation \eqref{eq:ode_p} yields our final governing equation, 
\begin{multline}
\label{eq:ode_p1}
\left(1+\frac{\phi_0 p(x)}{\gamma\bar{\phi}(x)}\right)^3 \left(1+\frac{\phi_0 p(x)}{\gamma\bar{\phi}(x)}+\frac{4b(x)}{\gamma L}\right)\frac{dp}{dx}\\ + 12\left(1+\frac{\phi_0 p(x)}{\gamma\bar{\phi}(x)}+\frac{b(x)}{\gamma L}\right) = 0,
\end{multline}
which captures all the physics of the system.
This equation, which is a first-order ODE for $p(x)$, is subject to the outlet boundary condition in equation \eqref{eq:outletp_nd}. Equation~\eqref{eq:ode_p1} is the key mathematical result that we employ below to formulate the forward and inverse problems for controlling the microchannel shape.

\begin{figure*}
\subfloat[]{\includegraphics[width=0.48\textwidth]{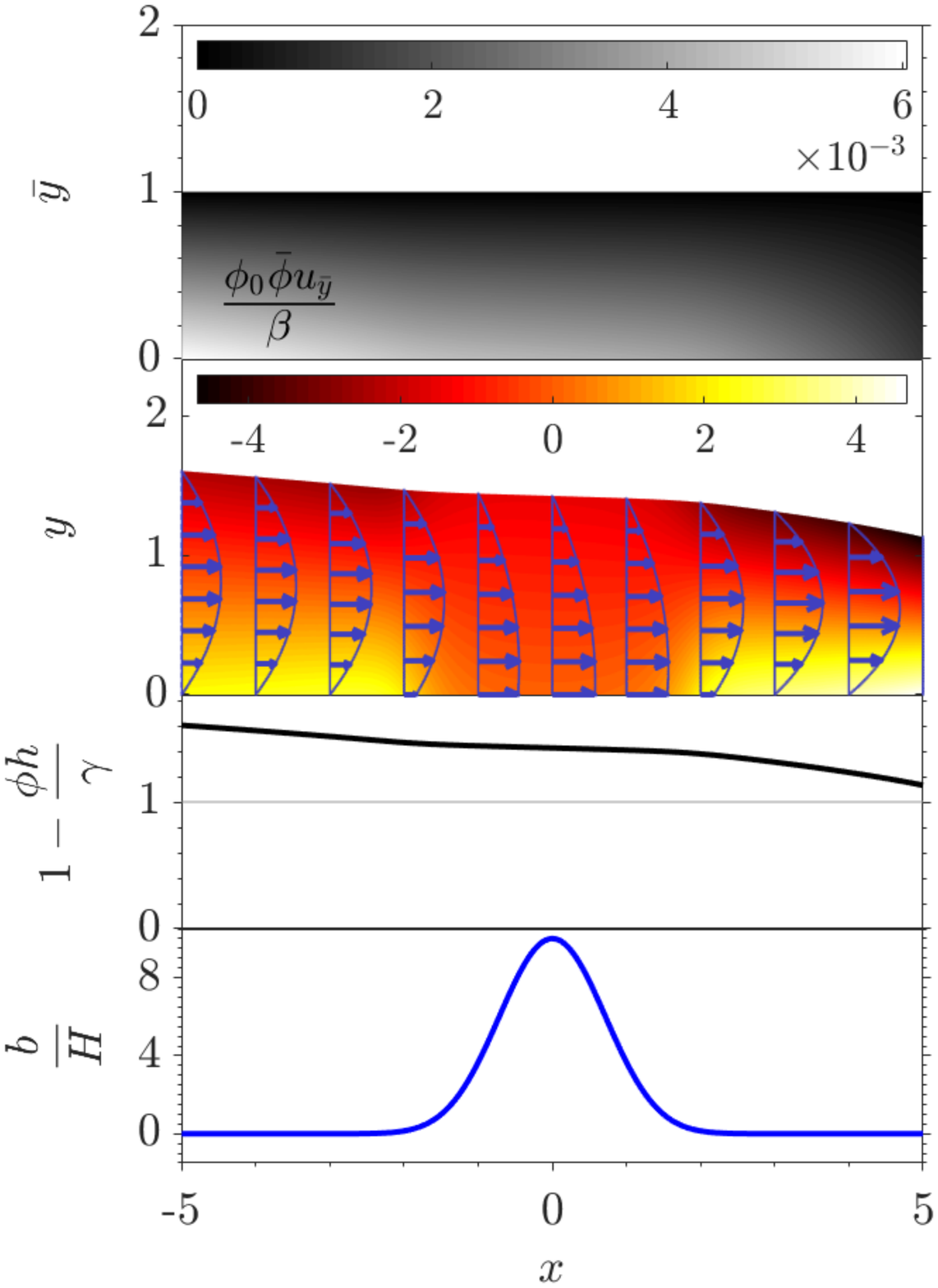}\label{subfig:slip_slow}}
\subfloat[]{\includegraphics[width=0.48\textwidth]{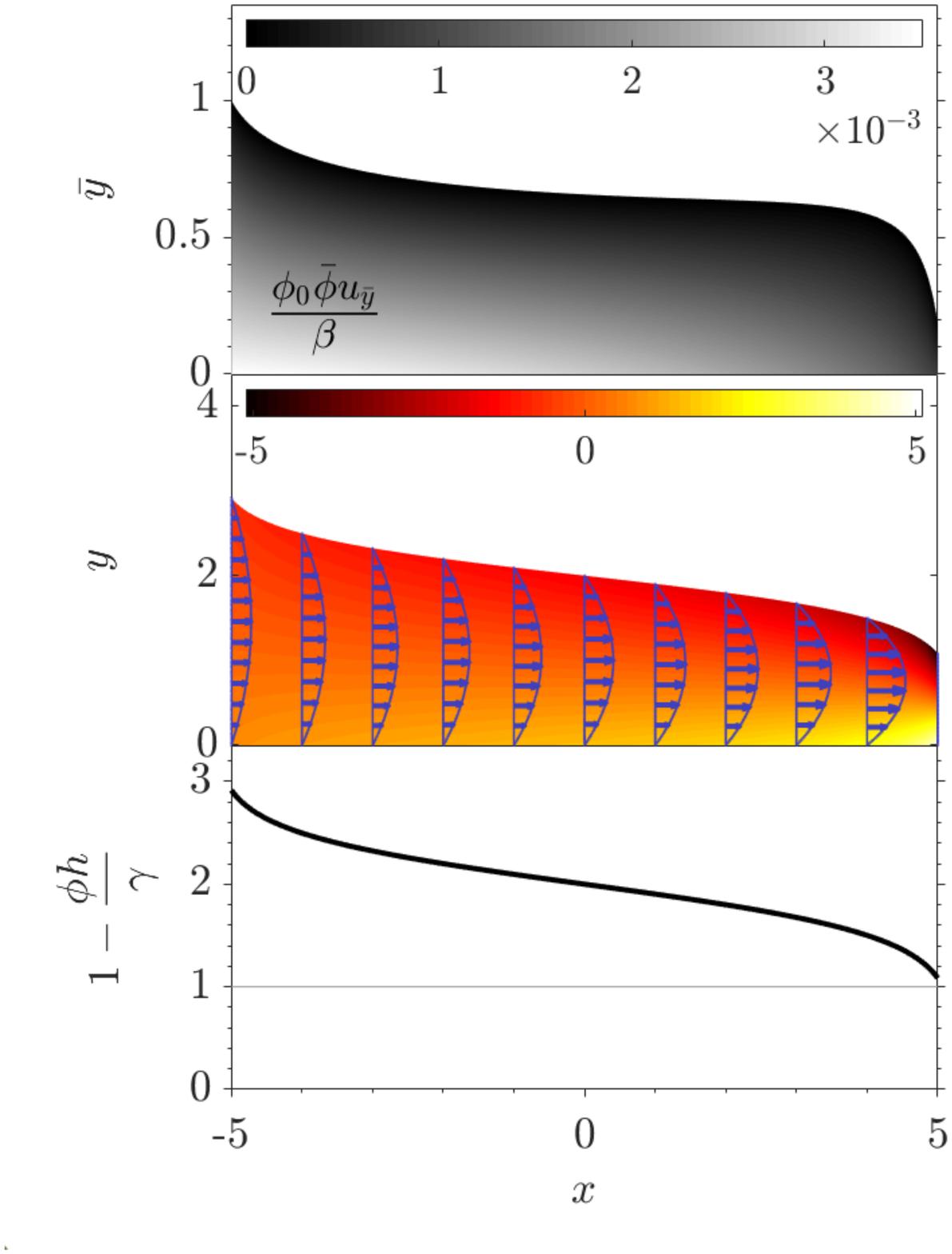}\label{subfig:taper_slow}}
\caption{(Color online.) The `slow-converging' deformed channel shape; description of individual panels is provided in the text. Here, (a) is obtained using slip length patterning, keeping the solid layer uniform constant: $\xi(x) = 1$; (b) is obtained using solid layer profiling without slip: $b(x)=0$. The dimensionless parameter values used are: $\gamma$ = $2\times 10^{-4}$, $\kappa$ = $0.2$, $\beta$ = $0.02$ (a) and $0.11$ (b), $\phi_0$ = $2.85\times 10^{-6}$ (a) and $1.53\times 10^{-5}$ (b), $\bar{p}_0$ = $35$ (a) and $5.9$ (b); the description of variables presented in the plots is available in the first paragraph of section \ref{subsec:canonical}.} 
\label{fig:slow}
\end{figure*}

\begin{figure*}
\subfloat[]{\includegraphics[width=0.48\textwidth]{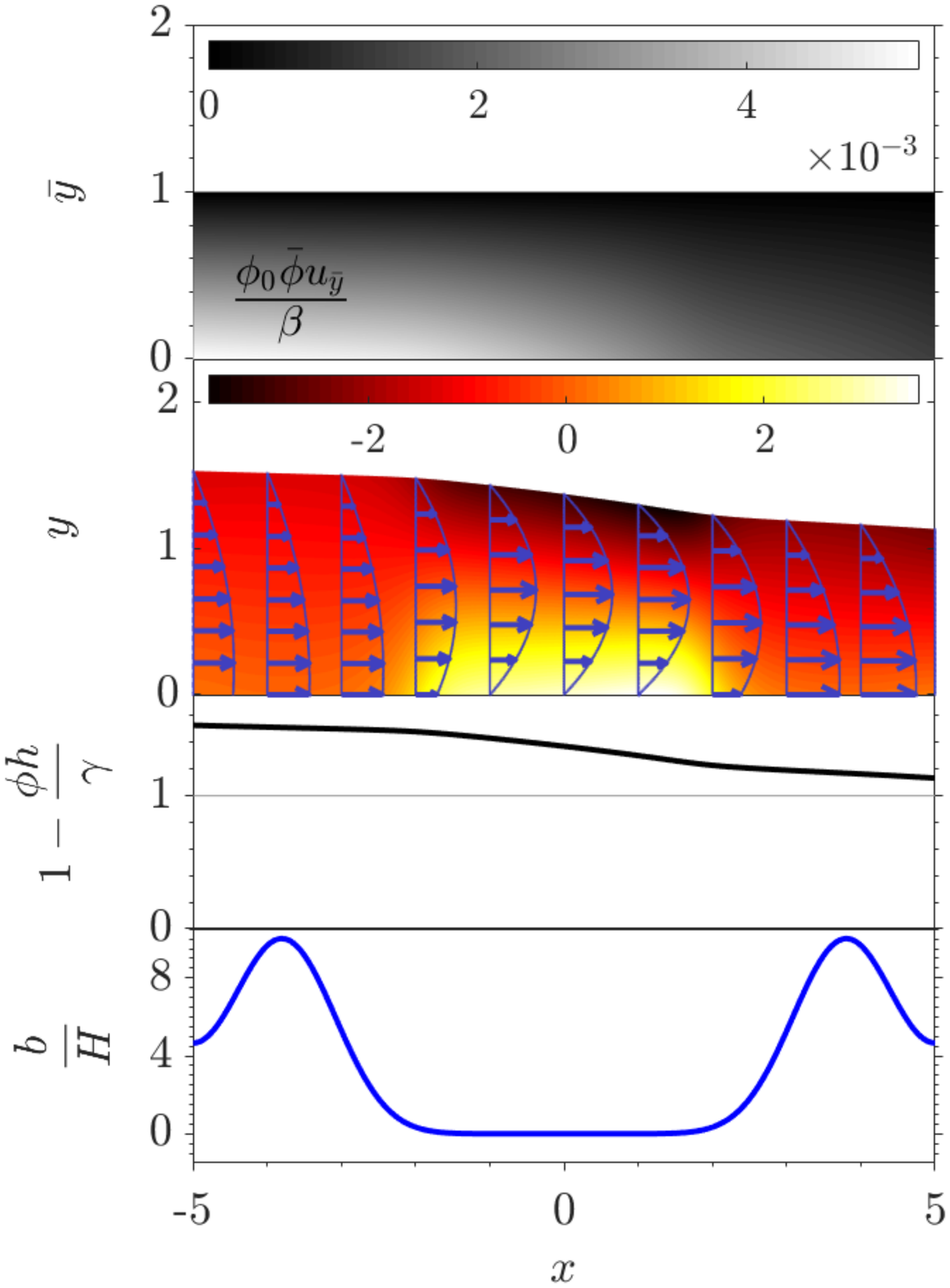}\label{subfig:slip_fast}}
\subfloat[]{\includegraphics[width=0.48\textwidth]{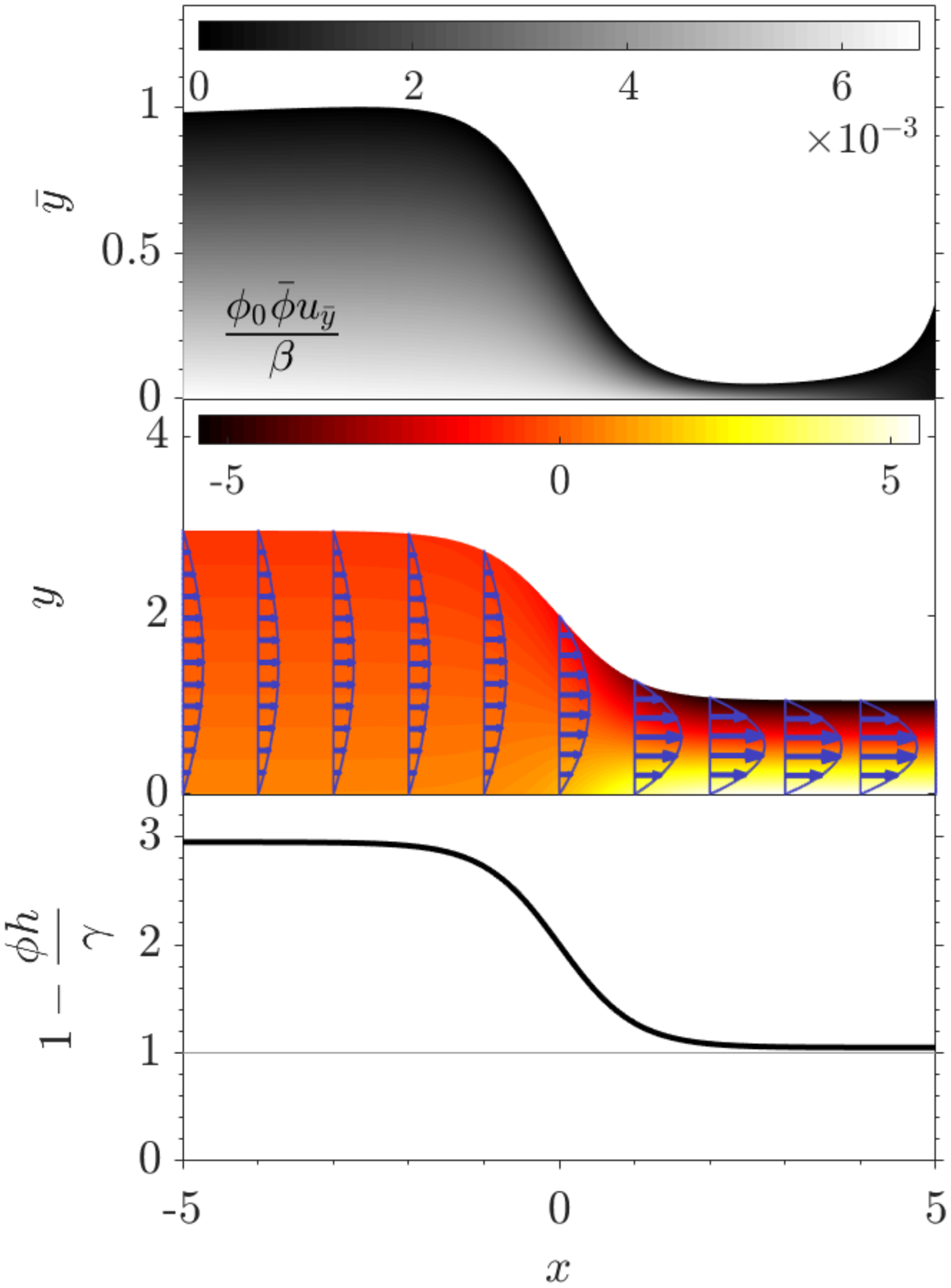}\label{subfig:taper_fast}}
\caption{(Color online.) The `fast-converging' deformed channel shape; description of individual panels is provided in the text. Here, (a) is obtained using slip length patterning, keeping the solid layer profiling uniform: $\xi(x) = 1$; (b) is obtained using solid layer profiling without slip: $b(x)=0$. The dimensionless parameter values used are: $\gamma$ = $2\times 10^{-4}$, $\kappa$ = $0.2$, $\beta$ = $0.02$ (a) and $0.06$ (b), $\phi_0$ = $2.85\times 10^{-6}$ (a) and $8.24\times 10^{-6}$ (b), $\bar{p}_0$ = $35$ (a) and $3.5$ (b); the description of variables presented in the plots is available in the first paragraph of section \ref{subsec:canonical}.}
\label{fig:fast}
\end{figure*}

\subsection{Inverse Problems and Solutions}\label{subsec:inverse}
We can define an inverse problem, `Inverse Problem A,' for which we are asked to find a function $b(x)$ that would yield a prescribed wall deflection shape $h=\bar{h}(x)$. To obtain a solution to this inverse problem, we first substitute $p$ from equation \eqref{eq:h} into  \eqref{eq:ode_p} and re-arrange to obtain
\begin{multline}
\label{eq:b_inverseprob}
b(x) = -\frac{\big[\gamma-\phi_0 \bar{h}(x)\big] L}{4} \\ \times \left\{\frac{12\gamma^3 - \big[\gamma-\phi_0 \bar{h}(x)\big]^3\frac{d }{d x}\big[\bar{\phi}(x) \bar{h}(x)\big]}{3\gamma^3 - \big[\gamma-\phi_0 \bar{h}(x)\big]^3\frac{d }{d x}\big[\bar{\phi}(x) \bar{h}(x)\big]}\right\}.
\end{multline}
Similarly, we substitute $p(x)$ from equation \eqref{eq:h} into equation \eqref{eq:outletp_nd} to get:
\begin{equation}
	\label{eq:outleth_nd}
	\bar{p}_0 = -\left[\bar{\phi}(x)\bar{h}(x)\right]_{x = {1}/{\kappa}},
\end{equation}
as the boundary condition. Equations~\eqref{eq:b_inverseprob} and \eqref{eq:outleth_nd} represent the first `inverse solution' obtained by our mathematical approach. Essentially, we compute what $b(x)$ and $\bar{p}_0$ (equivalently $p_0^*$, see equation \eqref{eq:outletp_nd}) should be imposed in order to obtain the desired deflection shape $h=\bar{h}(x)$, for a pre-set thickness variation $\bar{\phi}(x)$ (equivalently, a pre-set $\xi(x)$). 

We can also define another inverse problem, `Inverse Problem B,' in which we seek a $\bar{\phi}(x)$ (equivalently, a $\xi(x)$) that would yield a prescribed wall deflection shape $h=\bar{h}(x)$. To obtain a solution for this inverse problem, we first substitute $p(x)$ from equation \eqref{eq:h} into  \eqref{eq:ode_p} and re-arrange to obtain an ODE for $\bar{\phi}(x)$:
\begin{equation}
	\label{eq:ode_xi}
	\frac{d\left(\bar{\phi}\bar{h}\right)}{dx} = \frac {12 \left(1-\frac{\phi_0 \bar{h}(x)}{\gamma}+\frac{b(x)}{\gamma L}\right)}  {\left(1-\frac{\phi_0 \bar{h}(x)}{\gamma}+\frac{4b(x)}{\gamma L}\right) \left(1-\frac{\phi_0\bar{h}(x)}{\gamma}\right)^{3}},
\end{equation}
Equation \eqref{eq:ode_xi} is a first-order ODE in $\bar{\phi}(x)$, whose solution gives us $\xi(x)= {1}/{\bar{\phi}(x)}$. To solve equation \eqref{eq:ode_xi}, we require one boundary condition, which we take to be at the outlet, $\xi(1/\kappa) = 1$. After a solution is obtained, equation \eqref{eq:outleth_nd} gives us the outlet pressure.

It is well known that solutions to inverse problems may not be unique (or physically valid), often requiring a regularization. Thus, solving each of the proposed inverse problems requires some care to ensure a valid solution. Specifically, given a desired wall deflection shape, it is possible that the solution to equation \eqref{eq:b_inverseprob} or \eqref{eq:ode_xi} will yield negative values of $b$ or $\xi$, respectively, which is unphysical. Therefore, in the event that we obtain negative values, we offset the obtained solution up by the magnitude of the most negative value, which generates a new guess for the solution, now having a minimum value of zero. Now, however, the patterned slip length $b^*(x^*)$ can have values that are significantly higher than the undeformed channel height $H$. In such a case, we scale the obtained $b^*(x^*)$ so that the slip length maximizes to a value on the order of $10 H$. Similarly, we scale the solid layer profile $\xi(x)$ so that it has a maximum value of 1 along the channel length, i.e., $\xi(x) \mapsto \frac{1}{\max_x\xi(x)}\cdot\xi(x)$. We also re-scale $\beta \mapsto \textstyle{\max_x}\xi(x) \cdot \beta$ so that the dimensional solid layer profile is as close as possible to what is desired. This procedure ensures that we converge to a physically relevant solution to the inverse problem. Whenever these rescalings become necessary, it is indicative that the desired wall deflection $\bar{h}(x)$ is unrealistic. Nevertheless, in such situations, we still expect that the obtained shape (via the  regularization just explained) would be similar to the desired shape. 

For a setup in which we have utilized attuning of the solid layer profiling to obtain a desired wall deflection, we can additionally utilize the slip length patterning $b(x)$ to target a desired bottom wall shear rate axial variation. This amounts to another inverse problem: `Inverse Problem C', for $b(x)$, with $\bar{h}(x)$ already known. Evaluating equation \eqref{eq:shear} at the bottom wall, $y=0$, substituting ${dp}/{dx}$ from equation \eqref{eq:ode_p} into the latter, and performing further algebraic manipulations yields the sought-after expression for $b$ as:
\begin{equation}
	\label{eq:b_inverseshear_bottom}
	b(x) = -\frac{\gamma L}{4}\left(1-\frac{\phi_0h(x)}{\gamma}\right) \left[1-\frac{6({\partial v_x }/{\partial y})|_{y=0}}{\left(1-{\phi_0h(x)}/{\gamma}\right)^{2}}\right]^{-1},
\end{equation}
where $({\partial v_x }/{\partial y})|_{y=0}$ is the desired axial variation of the shear rate along the bottom wall of the fluid domain, which could be specified based on physiological considerations (for, say, cells) in a microfluidic experiment (as discussed in section~\ref{sec:intro}).

\subsection{Forward Problem and Numerical Scheme}\label{subsec:soln}

Equation \eqref{eq:ode_p1}, subject to the boundary condition in equation \eqref{eq:outletp_nd},  represents the `forward' problem mathematically. Once the solution for $p(x)$ is obtained, $h(x)$ is found from equation \eqref{eq:h}. Thus, to solve equations \eqref{eq:ode_p1} and \eqref{eq:outletp_nd}, we discretize the derivatives in equation \eqref{eq:ode_p1} using finite-differences, except at $x=1/\kappa$, where equation \eqref{eq:outletp_nd} is applied. We solve the resulting nonlinear algebraic system of equations using the multivariable Newton--Raphson method (described in appendix~\ref{sec:Newton}).

\begin{figure*}
\subfloat[]{\includegraphics[width=0.48\textwidth]{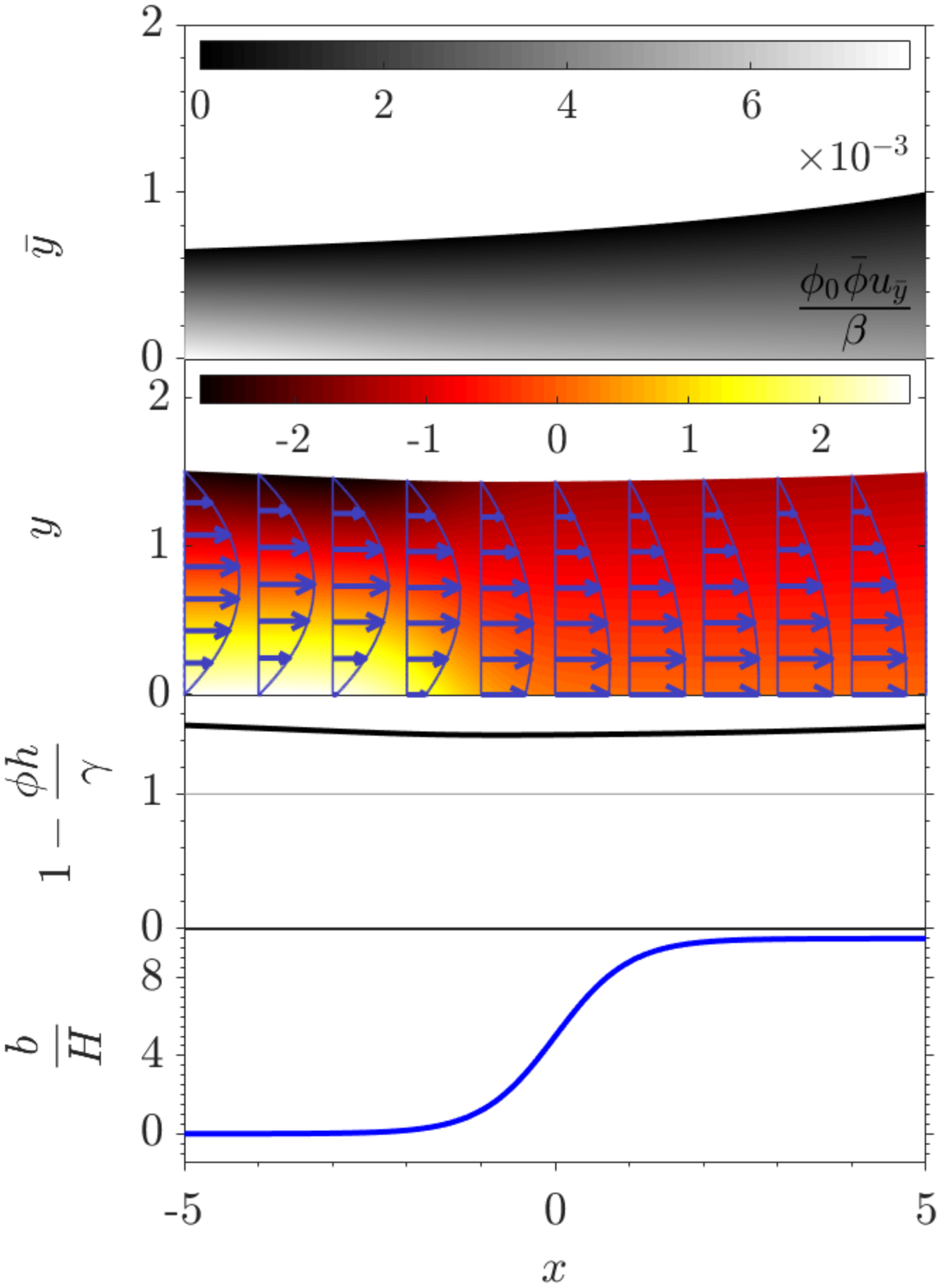}\label{subfig:slip_dip}}
\subfloat[]{\includegraphics[width=0.48\textwidth]{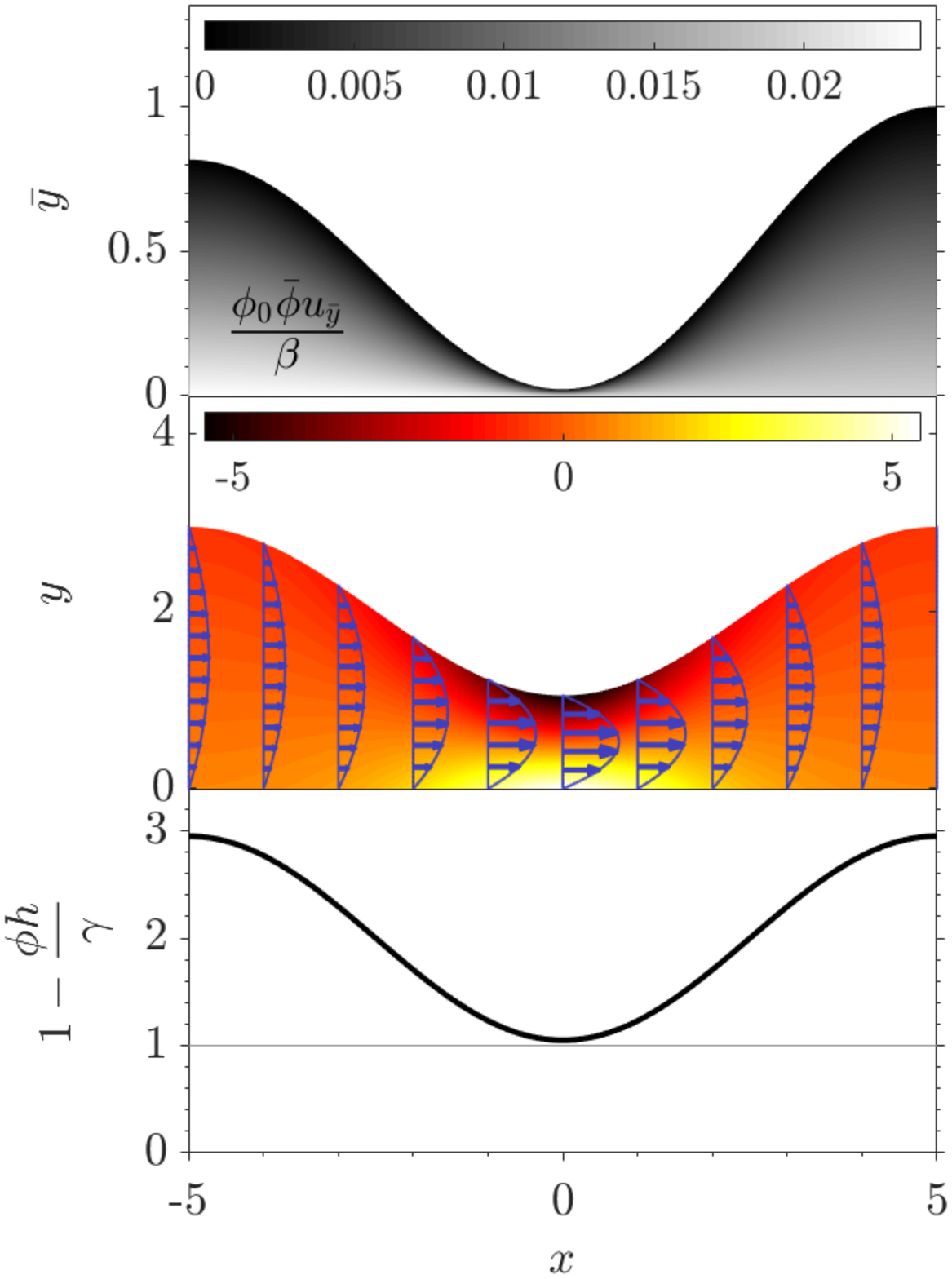}\label{subfig:taper_dip}}
\caption{(Color online.) The converging-diverging deformed channel shape; description of individual panels is provided in the text. Here, (a) is obtained using slip length patterning aided by solid layer profiling, and (b) is obtained using solid layer profiling without slip. The dimensionless parameter values used are: $\gamma$ = $2\times 10^{-4}$, $\kappa$ = $0.2$, $\beta$ = $0.02$ (a) and $0.02$ (b), $\phi_0$ = $2.85\times 10^{-6}$ (a) and $2.85\times 10^{-6}$ (b), $\bar{p}_0$ = $35$ (a) and $137.0$ (b); the description of variables presented in the plots is available in the first paragraph of section \ref{subsec:canonical}.}
\label{fig:dip}
\end{figure*}

\begin{figure*}
\subfloat[]{\includegraphics[width=0.48\textwidth]{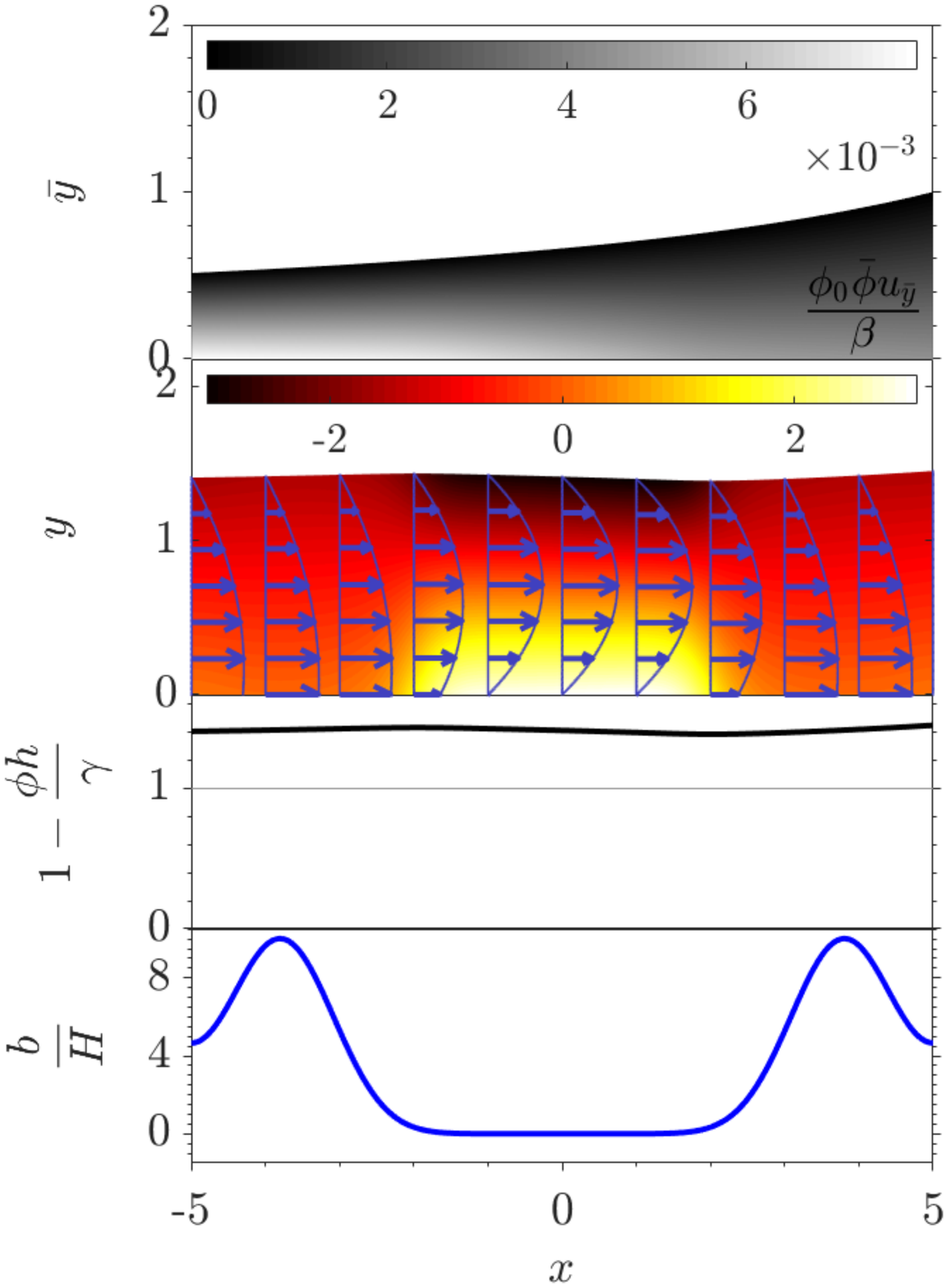}\label{subfig:slip_wave}}
\subfloat[]{\includegraphics[width=0.48\textwidth]{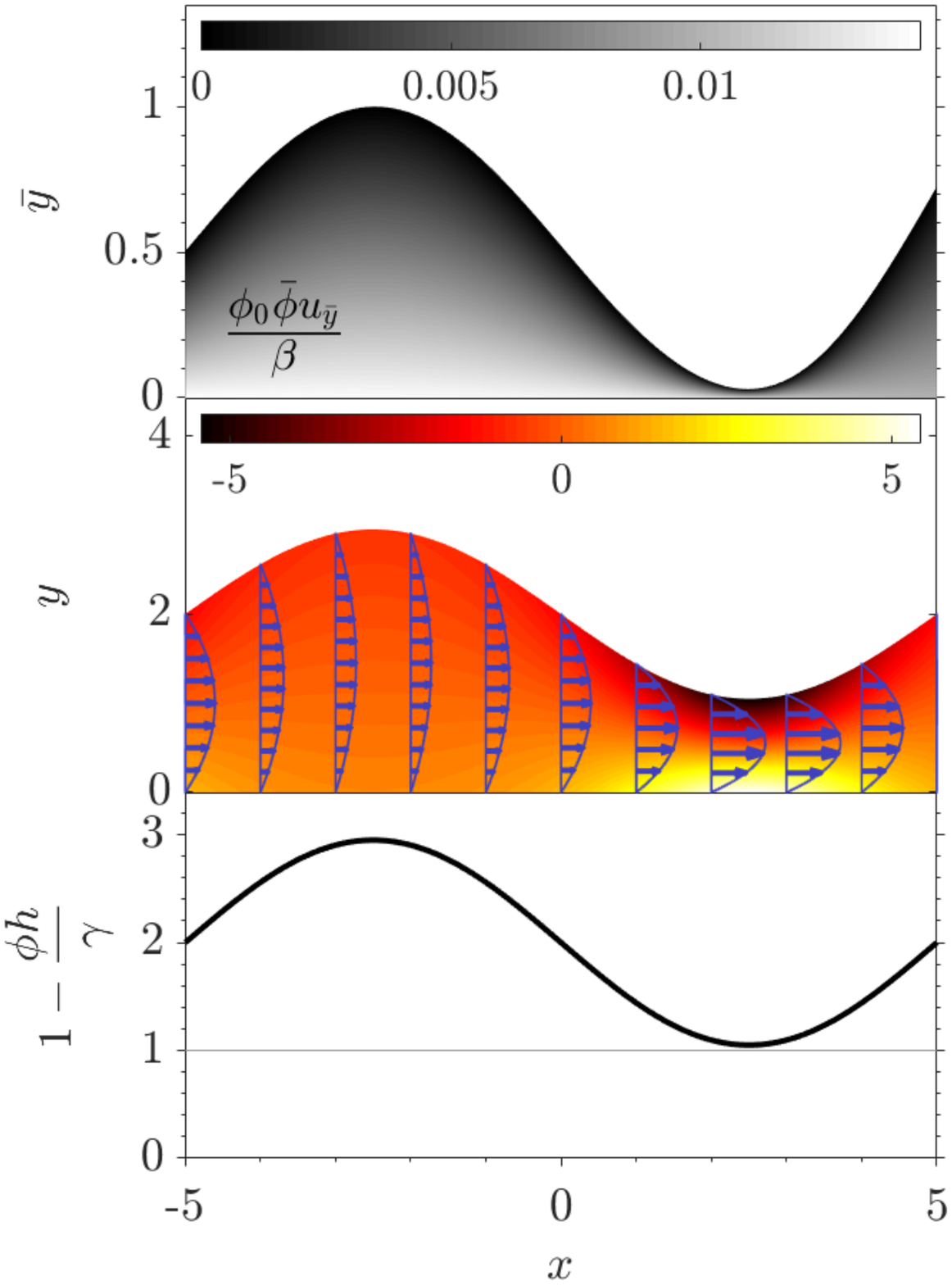}\label{subfig:taper_wave}}
\caption{(Color online.) The diverging--converging-diverging deformed channel shape; description of individual panels is provided in the main text. Here, (a) is obtained using slip length patterning aided by solid layer profiling, and (b) is obtained using solid layer profiling without slip. The dimensionless parameter values used are: $\gamma$ = $2\times 10^{-4}$, $\kappa$ = $0.2$, $\beta$ = $0.02$ (a) and $0.03$ (b), $\phi_0$ = $2.85\times 10^{-6}$ (a) and $3.90\times 10^{-6}$ (b), $\bar{p}_0$ = $31.6$ (a) and $70.3$ (b); the description of variables presented in the plots is available in the first paragraph of section \ref{subsec:canonical}.}
\label{fig:wave}
\end{figure*}

\begin{figure*}[!thb]
\subfloat[]{\includegraphics[width=0.48\textwidth]{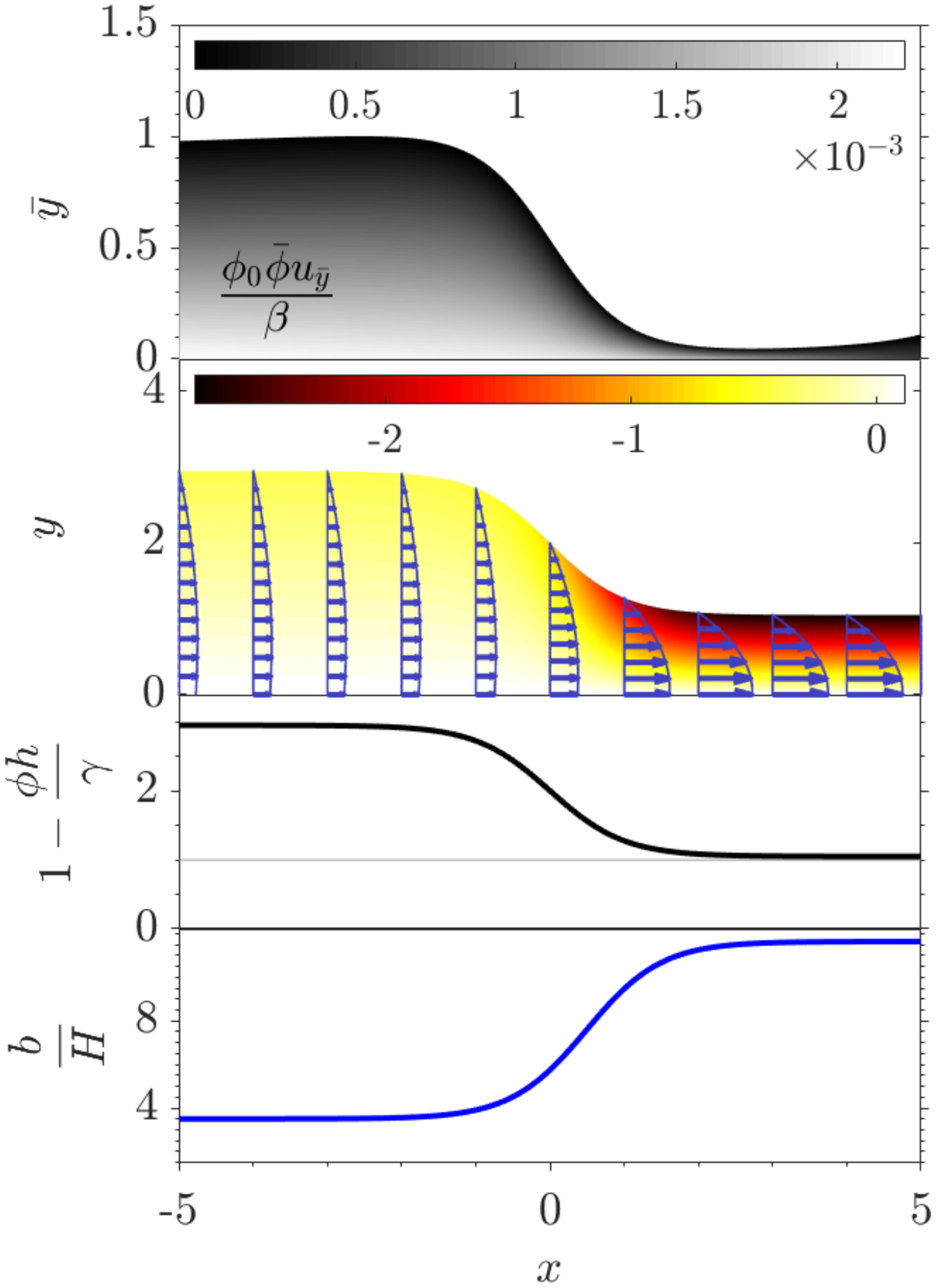}\label{subfig:taper_fast_sh}}
\subfloat[]{\includegraphics[width=0.48\textwidth]{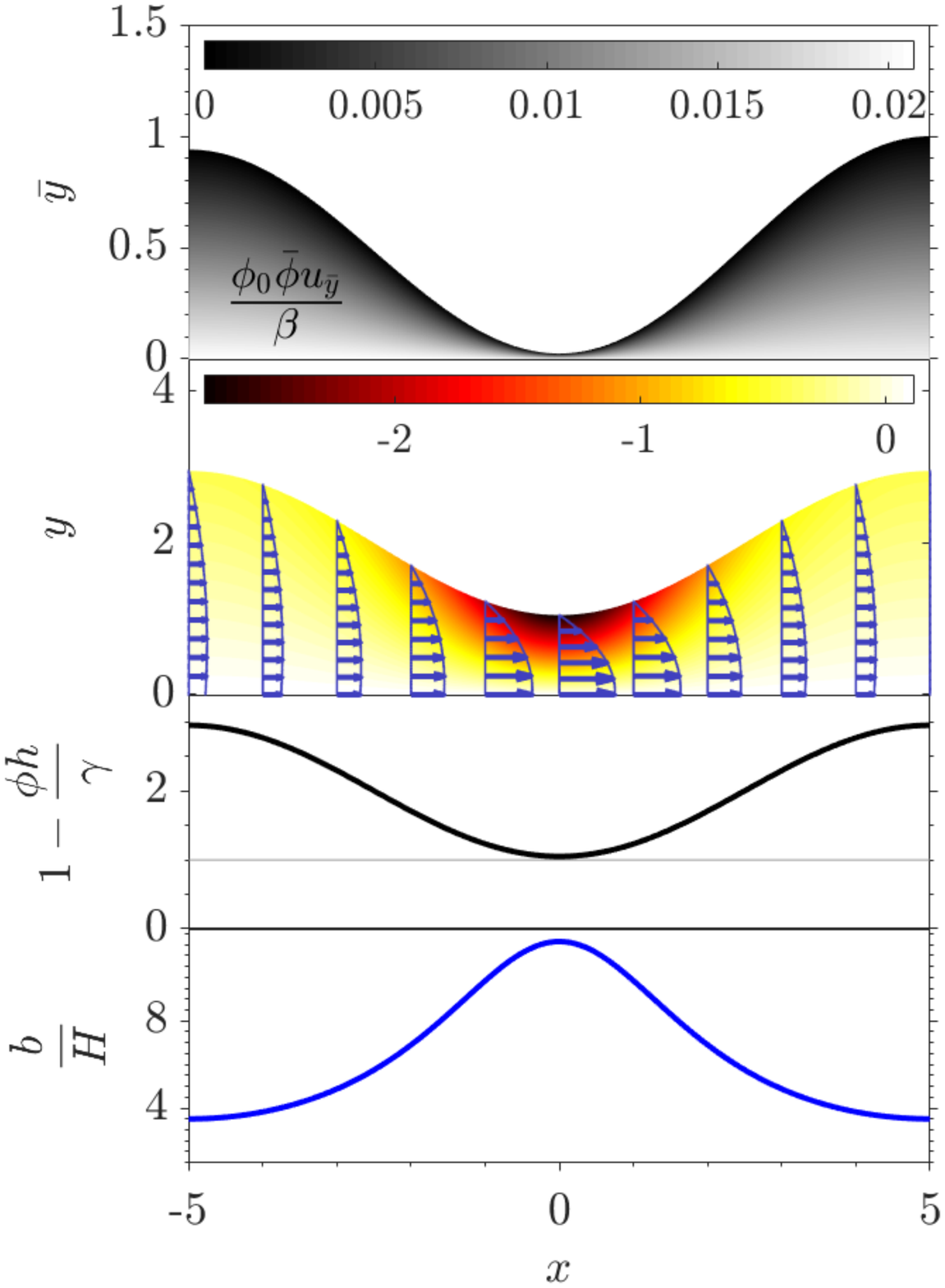}\label{subfig:taper_dip_sh}}
\caption{(Color online.) Analogous solutions of (a) fast-converging (figure \protect\subref*{subfig:taper_fast}) and (b) converging-diverging (figure \protect\subref*{subfig:taper_dip}) shapes, now using slip length patterning to maintain a low shear along the rigid bottom wall of the fluid domain; description of individual panels is provided in the main text The dimensionless parameter values used are: $\gamma$ = $2\times 10^{-4}$, $\kappa$ = $0.2$, $\beta$ = $0.18$ (a) and $0.02$ (b), $\phi_0$ = $2.60\times 10^{-5}$ (a) and $2.85\times 10^{-5}$ (b), $\bar{p}_0$ = $3.56$ (a) and $137$ (b); the description of variables presented in the plots is available in the first paragraph of section \ref{subsec:canonical}.}
\label{fig:shear}
\end{figure*}

\begin{figure*}[!thb]
\centering
\subfloat[]{\includegraphics[height=0.33\textwidth]{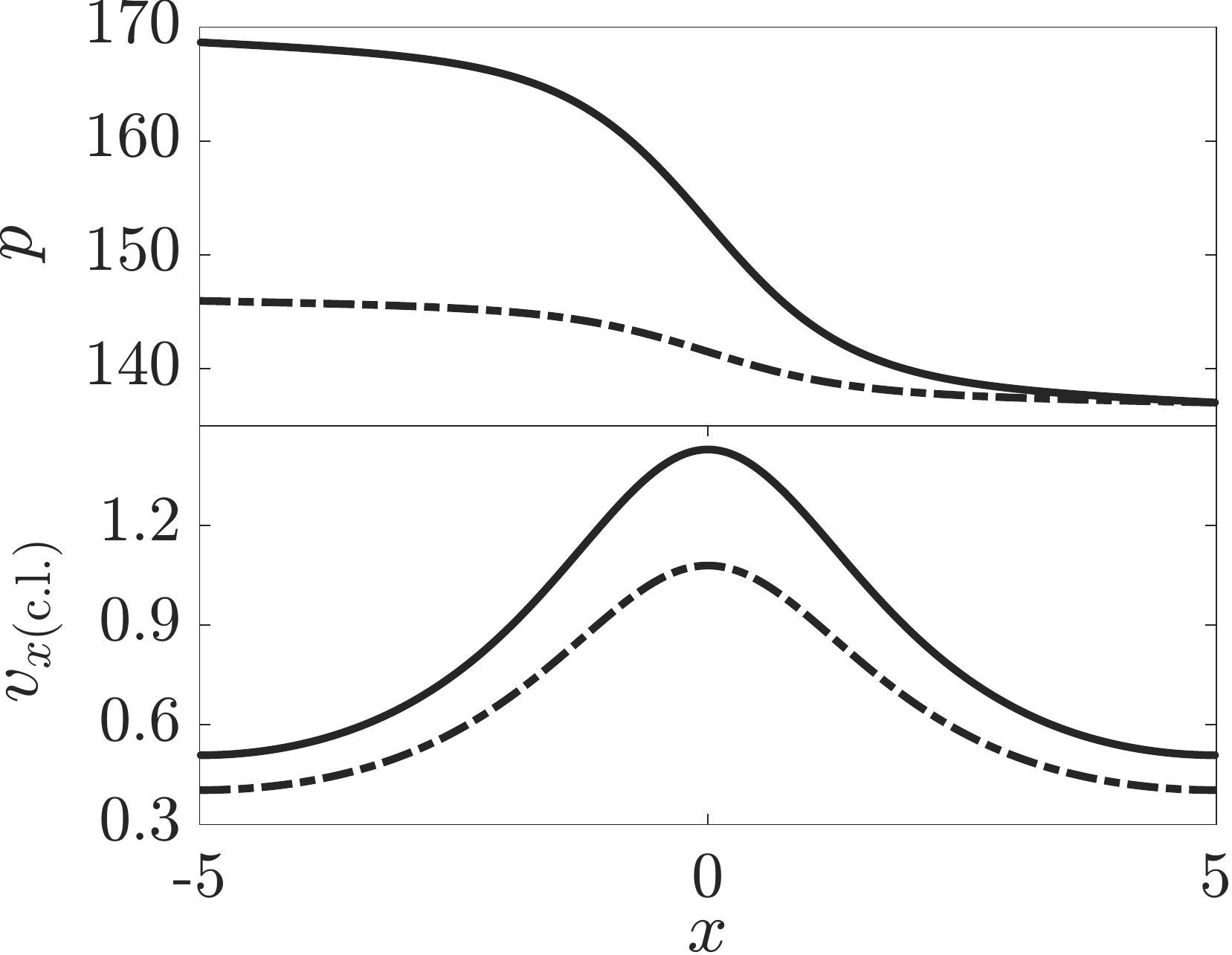}\label{subfig:dip_vx_p_x}}\hspace{15pt}
\subfloat[]{\includegraphics[height=0.33\textwidth]{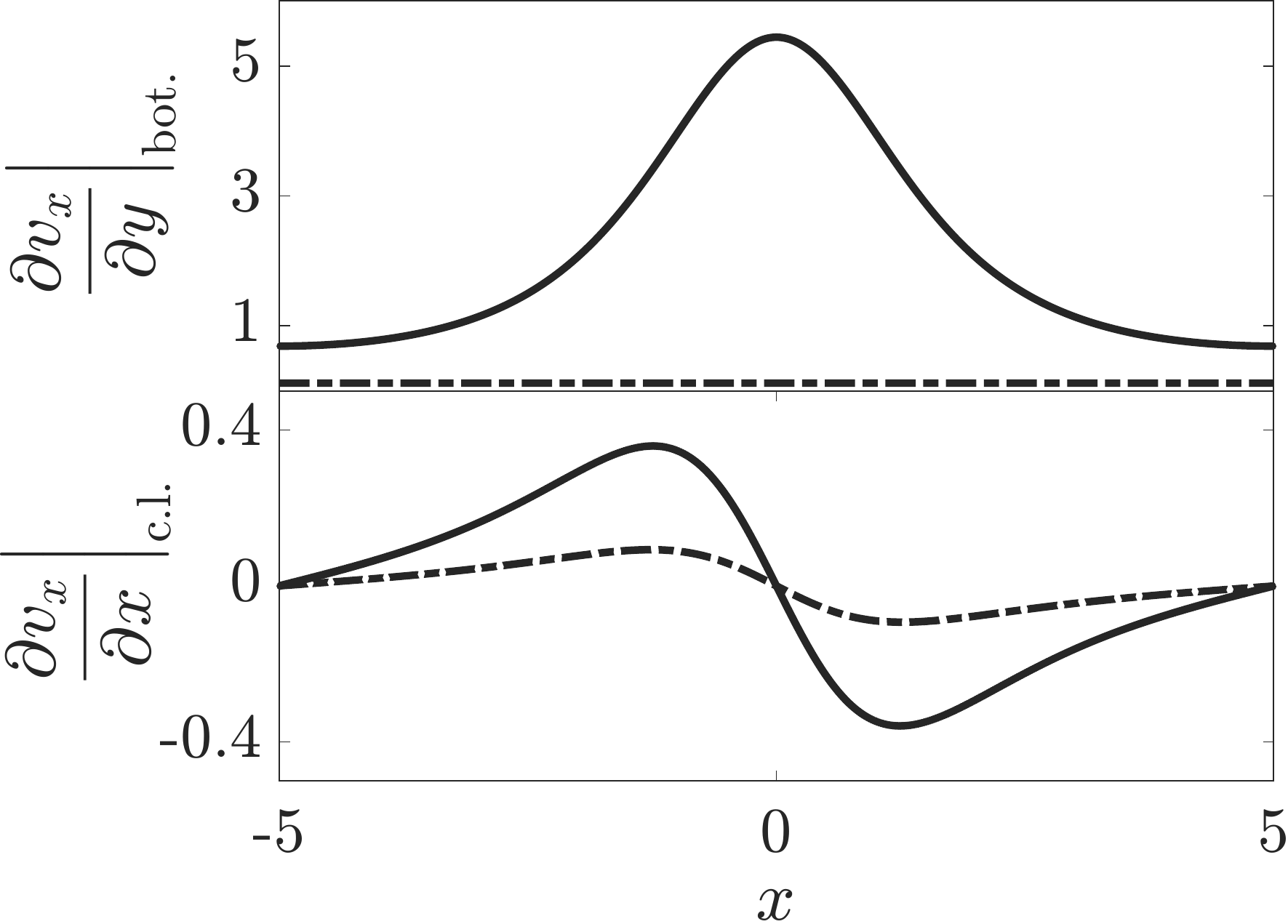}\label{subfig:dip_dvxdx_dvxdy_x}}
\caption{(Color online.) Axial variation of (a) pressure (top) and axial velocity at $y = (1-{\phi h/\gamma})/2$ (bottom), and, (b) flow shear rate at $y = 0$ (top panel) and flow extensional rate at $y = (1-{\phi h/\gamma})/2$ (bottom panel), for the converging-diverging channel without slip (corresponding to figure \protect\subref*{subfig:taper_dip}, represented by solid curves here) and the converging-diverging channel with bottom wall slip length patterning to reduce the shear rate (corresponding to figure \protect\subref*{subfig:taper_dip_sh}, represented by dashed-dotted curves here); the subscript `c.l.' stands for central-line and the subscript `bot.' stands for bottom wall. Note that the central line, represented by $y = (1-{\phi h/\gamma})/2$, is actually not a straight line but a curve because of the top wall deformation.}
\label{fig:dip_ext}
\end{figure*}


\section{Results}\label{sec:results}

With the theory formulated in section \ref{sec:math} in hand, our objective now is to achieve desired shapes for the microchannel, i.e., to obtain pre-determined fluid--solid interface deflection shapes during flow. We demonstrate four types of wall shapes. We first discuss, in section \ref{subsec:canonical}, these general categories of geometric shape variations, specifically presenting two versions of each. For each set, the first shape is obtained predominantly using slip length patterning by solving Inverse Problem A, where, either we do not use any solid layer profiling (figures \subref*{subfig:slip_slow} and \subref*{subfig:slip_fast} ahead) or we use simplistic solid layer profiling to complement the slip length patterning as slip length patterning alone falls short in recovering the desired channel shape (figures \subref*{subfig:slip_dip} and \subref*{subfig:slip_wave} ahead). The second shape for each set is an exaggerated version of the first shape, which is obtained by solely attuning the solid layer profiling (by solving Inverse Problem B), without considering any slip. Thus, we show that attunement of the solid layer profiling is a significantly more effective tool for controlling the fluid--solid interface's shape.
Subsequently, in section \ref{subsec:extent}, we present the consequence of the extent of implementation of the obtained slip length patterning (by solving Inverse Problem A) and the obtained solid layer profiling (by solving Inverse Problem B) on the recovered channel shape under flow, for one of the canonical shapes. 
Lastly, in section \ref{subsec:pdrop}, we discuss the necessary pressure drop across the channel required to maintain a given flow rate. 

\subsection{Canonical Channel Shapes}\label{subsec:canonical}

The four sets of canonical shapes are presented in and discussed using figures \ref{fig:slow} to \ref{fig:wave}. For each set, the solutions obtained using slip length patterning are presented in subfigures (a) (on the left), while those obtained by using solid layer profiling are presented in subfigures (b) (on the right). In each subfigure (a), we plot four quantities. The bottom-most plot shows the patterned slip-length made dimensionless by the undeformed channel height, i.e., the variation of ${b(x)}/{H}$ with $x$. The curve second from the bottom is the dimensionless deformed height of the channel, i.e., the variation of $1-{\phi_0h(x)}/{\gamma}$ with $x$. For reference, the top channel wall in the absence of any deformation is presented as the thin grey line. In the plot third from bottom, we present a `heat map' of the dimensionless shear rate in the deformed fluid domain, i.e., a heat map of ${\partial v_x}/{\partial y}$ in the deformed fluid domain. On top of this heat map, we have also presented the fluid velocity profile using the arrows as depicted. Finally, in the topmost plots, we show a greyscale heat map of the ratio of $\bar{y}$-deformation referenced to local solid layer thickness, i.e. the ratio $\displaystyle \frac{u_{\bar{y}^*(x^*,\bar{y}^*)}}{\Delta^*(x^*)}$, whose dimensionless version reads $\displaystyle \frac{\phi_0\bar{\phi}u_{\bar{y}}}{\beta}$. We emphasize that this heat map, which has been presented in the deformed solid domain, remains identical in the undeformed solid domain (not presented), a consequence of being restricted to the infinitesimal-strain regime of deformation. Each of subfigures (b) shows three plots, which convey the same variations as in the second, third and fourth plots from the bottom in subfigure (a). Lastly, in figure \ref{fig:shear}, variants for two of the canonical shapes have been presented (where the solution has been obtained using Inverse Problem C), and the scheme of plots for both subfigures of figure \ref{fig:shear} is the same as subfigures (a) of figures \ref{fig:slow} to \ref{fig:wave}.

The first two example shapes, which are converging along the channel length and `slow--converging' near the middle of the channel, are presented in figure \ref{fig:slow}. The shape in figure~\subref*{subfig:slip_slow} is obtained using slip length patterning, keeping the solid layer profiling uniform: $\xi(x) = 1$. This shape exhibits a slower variation of the deformed top wall of the channel near the center, with faster variations near the inlet and outlet. This variation is accomplished by a patch of slip flow near the centre of the channel, with no-slip for the rest. The central region of the fluid domain exhibits significantly lower shear rate due to slip, with the shear rate being large on either side. The shape presented in figure \subref*{subfig:taper_slow} is obtained by solid layer profiling without considering slip. Now, we observe significant deflection of the upper boundary of the fluid domain, unlike in figure~\subref*{subfig:slip_slow}. Similarly, the variations near the inlet and the outlet are more pronounced in figure \subref*{subfig:taper_slow} than in figure \subref*{subfig:slip_slow}.

The next two example shapes, which are converging along the channel length and `fast--converging' near the middle of the channel, are presented in figure \ref{fig:fast}. The shape in figure~\subref*{subfig:slip_fast} is obtained using slip length patterning, keeping the solid layer profiling uniform: $\xi(x) = 1$. This shape exhibits faster variation of the deformed top wall of the channel near the center, with slower variations near the inlet and outlet. This variation is accomplished by having two patches of slip near the inlet and the outlet of the channel, with no-slip in the middle. The central region of the fluid domain exhibits a high shear rate in the rapidly converging channel shape. The shear rate is lower on either side, due to slip there. The shape presented in figure~\subref*{subfig:taper_fast} is obtained using solid layer profiling without considering slip. Like the slow--converging example above, we observe significant deflection of the fluid domain's top boundary in figure \subref*{subfig:taper_fast} compared to figure \subref*{subfig:slip_fast}. 

The next two example shapes generated by our passive control strategy, which are `converging-diverging' shapes, are presented in figure \ref{fig:dip}. The shape in figure~\subref*{subfig:slip_dip} is obtained using slip length patterning with simplistic solid layer profiling to complement. At first, a solution for $\xi(x)$ with no-slip is obtained such that when subjected to flow, the fluid domain's height bulges and assumes a linearly-convergent shape, i.e., $1-{\phi_0h(x)}/{\gamma}$ decreases linearly from $1.62$ at the inlet to $1.5$ at the outlet. Subsequently, upon applying slip length patterning as presented in the bottom panel of figure \subref*{subfig:slip_dip}, the shape exhibits a dip near the channel center, i.e., the deformed fluid domain has a converging-diverging shape. This is accomplished by having no-slip for the first half of the channel length and enhanced slip for the second half. The channel exhibits a converging shape with high shear rate near the inlet, and significantly lower shear rate and a diverging shape near the outlet. The shape presented in figure~\subref*{subfig:taper_dip} is obtained using solid layer profiling without considering slip. Like the former two sets, we observe significant deflection of the fluid domain's top boundary. Furthermore, the fluid domain (channel) is practically undeformed near the centre ($x=0$). Nevertheless, we emphasize that the channel is overall bulged in both figures~\subref*{subfig:slip_dip} and \subref*{subfig:taper_dip}, i.e., the deformed top wall position at any point is higher than the undeformed state.

The next two example shapes, which have the `diverging-converging-diverging' shape, are generated by our passive control strategy and are presented in figure \ref{fig:wave}. The shape in figure~\subref*{subfig:slip_wave} is obtained using slip length patterning with simplistic solid layer profiling to complement. At first, a solution for $\xi(x)$ with no-slip is obtained such that when subjected to flow, the fluid domain's height bulges and assumes a linearly-convergent shape, i.e., $1-{\phi_0h(x)}/{\gamma}$ decreases linearly from $1.5$ at the inlet to $1.45$ at the outlet. Subsequently, we apply two patches with slip near the inlet and the outlet of the channel, with no-slip in the middle (i.e. the slip length patterning along the bottom panel of figure \subref*{subfig:slip_wave}, which is very similar to the slip length patterning in the bottom panel of figure \subref*{subfig:slip_fast}). With this slip length patterning implemented, the channel shape we obtain under flow is a wavy one. It is evident that the `rotation' of the deflected wall shape from the fast-converging of figure \ref{fig:fast} to the diverging-converging-diverging one here is the outcome of attuning the solid layer profiling. The shape presented in \subref*{subfig:taper_wave} is obtained using solid layer profiling without slip. As with the three example sets of shapes above, we observe significant deflection of the fluid domain's top boundary. Furthermore, the fluid domain (channel) is practically undeformed at $x=2.5$. Again, we emphasize that the channel is overall bulged for both (a) and (b).

Two features are common to the four sets of example shapes presented above. First, for each of the shapes presented in (b), the shape that is its mirror image about $x=0$ is also obtainable using solid layer profiling without considering slip. Second, when we apply the slip length patterning of (a) on the solution from (b), the shear rate variation in (a) is superimposed onto the shear-rate variation in (b), without significant change in the top wall shape. This observation leads us to conclude that the slip length patterning is a weak mechanism when it comes to controlling the channel shape under flow, but it is effective for controlling the shear rate in the flow.

Next, we present two shapes that have been obtained using solid layer profiling without considering slip, and subsequently, slip length patterning has been obtained (by solving Inverse Problem C) to target a desired axial variation of bottom wall shear rate. Specifically, we obtain analogues of the fast-converging (figure \subref*{subfig:taper_fast}) and converging-diverging (figure \subref*{subfig:taper_dip}) setups for this purpose. These analogues are presented in figure \ref{fig:shear}. The deformed channel shape for each analogue is the same as the original. However, we have applied slip length patterning for each, such that the shear rate at the bottom wall has a low magnitude, as well as a smaller gradient---observe the bottom region (at and near $y=0$) of the heat map of the shear rate, presented in the second panel from top in figures \subref*{subfig:taper_fast_sh} and \subref*{subfig:taper_dip_sh} each. 

Lastly, in figure \ref{fig:dip_ext}, we present some key aspects of the flow in the converging-diverging channel. The solid curves correspond to no slip at bottom wall (corresponding to figure \subref*{subfig:taper_dip}) and the dashed-dotted curves correspond to slip patterning at bottom wall targeted at arresting the flow shear rate (corresponding to figure \subref*{subfig:taper_dip_sh}). The channel shape control is done by solid layer profiling for both line types. The axial flow velocity and the flow extensional rate as presented in the bottom panels of figure \ref{fig:dip_ext}, for either of the slip-patterning situations, exhibit appreciable qualitative similarity to the targeted variations of these variables by \citet{Zografos2016} in their study on designing of  converging-diverging microchannels. Given the non-dimensional nature of our analysis, appreciable quantitative similarity is also expected to be obtainable on a per-application basis, with a suitable choice of materials, geometry and flow conditions. Additionally, it can be observed that, as we introduce slip to arrest bottom wall shear rate (transitioning from solid curves to dashed-dot curves), the axial velocity and the flow extensional rate, although decreased in magnitude, retain their general form. However, the shear rate transitions from being finite to infinitesimal. This indicates that the design of a converging-diverging channel as presented in figure \subref*{subfig:taper_dip_sh} can be useful for applications in which a hyperbolic flow behavior is desired near the channel center, even while we may want the shear rate at the bottom wall to be arrested in magnitude.

\begin{figure*}[!thb]
\centering
\subfloat[]{\includegraphics[width=0.4\textwidth]{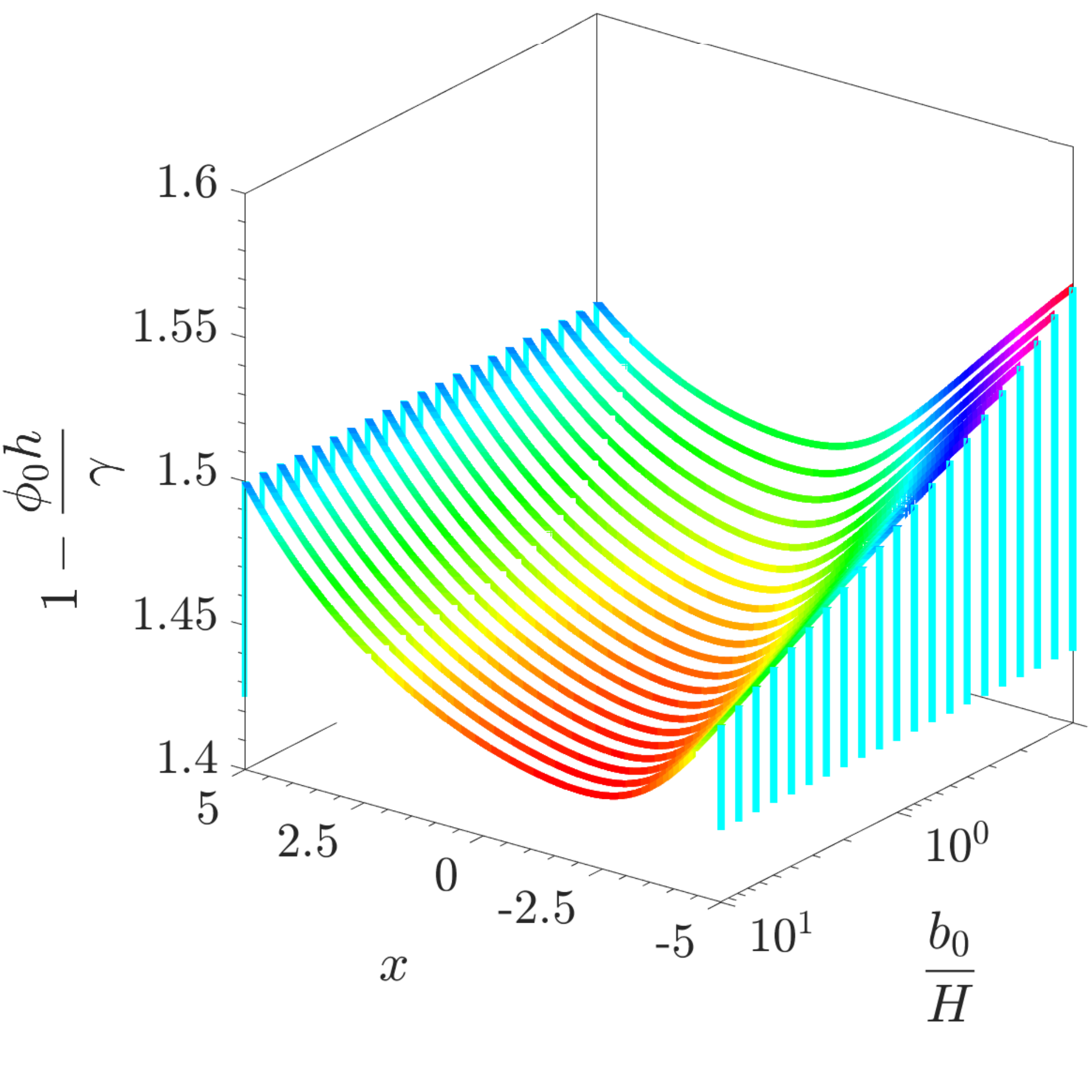}\label{subfig:sweep_slip}}\hspace{15pt}
\subfloat[]{\includegraphics[width=0.4\textwidth]{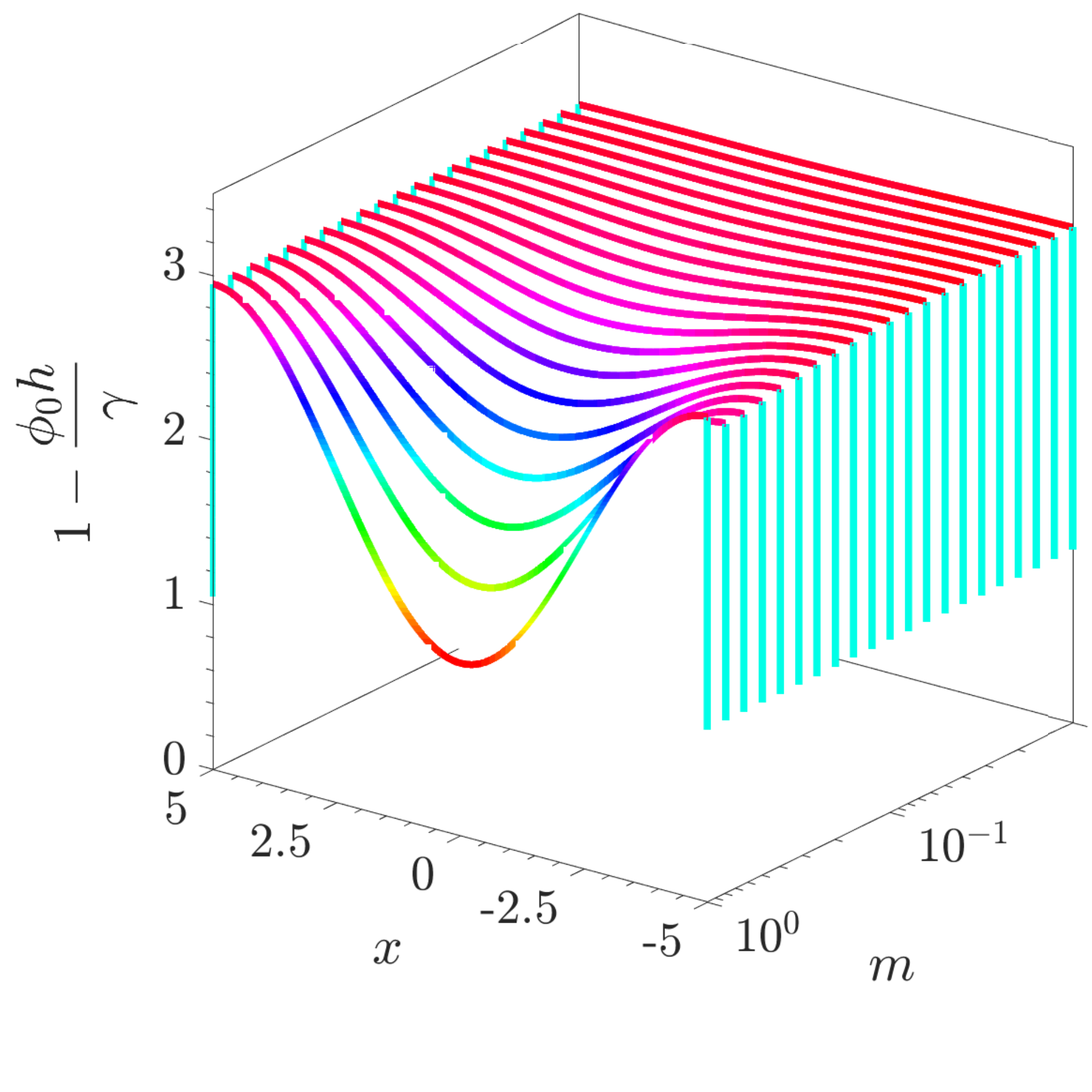}\label{subfig:sweep_taper}}
\caption{(Color online.) Parametric variation (note the logarithmic scale) of the dimensionless deformation shape of the fluid domain's top boundary with (a) slip length patterning (controlled via $b_0/H$) and (b) via solid layer profiling (controlled via $m$), for the converging-diverging case (figure \ref{fig:dip}).}
\label{fig:sweep_taper}
\end{figure*}

\begin{figure*}[!htb]
\centering
\begin{tabular}{c@{\hskip 10pt}l@{\hskip 10pt}c@{\hskip 10pt}l}	
\hline\\
\textbf{Color}														&	\textbf{Channel shape}																&	
\textbf{Linetype}													&	\textbf{Solution category}															\\[3pt]
\begin{tikzpicture}\fill[red] (0,0) circle (2.5pt);\end{tikzpicture}					&	slow-converging (figure \ref{fig:slow})												&	
\begin{tikzpicture}\draw[line width = 2pt] (0,0) -- (1,0);\end{tikzpicture}				&	undeformable top wall (rigid solid layer)											\\[3pt]
\begin{tikzpicture}\fill[blue] (0,0) circle (2.5pt);\end{tikzpicture}				&	fast-converging (figures \ref{fig:fast} and \subref*{subfig:taper_fast_sh})				&	
\begin{tikzpicture}\draw[line width = 2pt, dash dot, dash pattern={on 5pt off 2pt on 2pt off 2pt}] (0,0) -- (1,0);\end{tikzpicture}			&	actual solution (as presented in the respective figures)							\\[3pt]
\begin{tikzpicture}\fill[green] (0,0) circle (2.5pt);\end{tikzpicture}				&	converging-diverging (figures \ref{fig:dip} and \subref*{subfig:taper_dip_sh})			&	
\begin{tikzpicture}\draw[line width = 1pt, dash dot, dash pattern={on 5pt off 2pt on 2pt off 2pt}] (0,0) -- (1,0);\end{tikzpicture}&	constant thickness solid (replacing $\xi$ by $\tfrac{1}{\kappa}\int_{-1/\kappa}^{1/\kappa}\xi(x)\,dx$)															\\[3pt]
\begin{tikzpicture}\fill[magenta] (0,0) circle (2.5pt);\end{tikzpicture}				&	diverging-converging-diverging (figure \ref{fig:wave})								&	
~																	&	~	\\[3pt]
\hline
\end{tabular}
\subfloat[]{\includegraphics[width=0.3\textwidth]{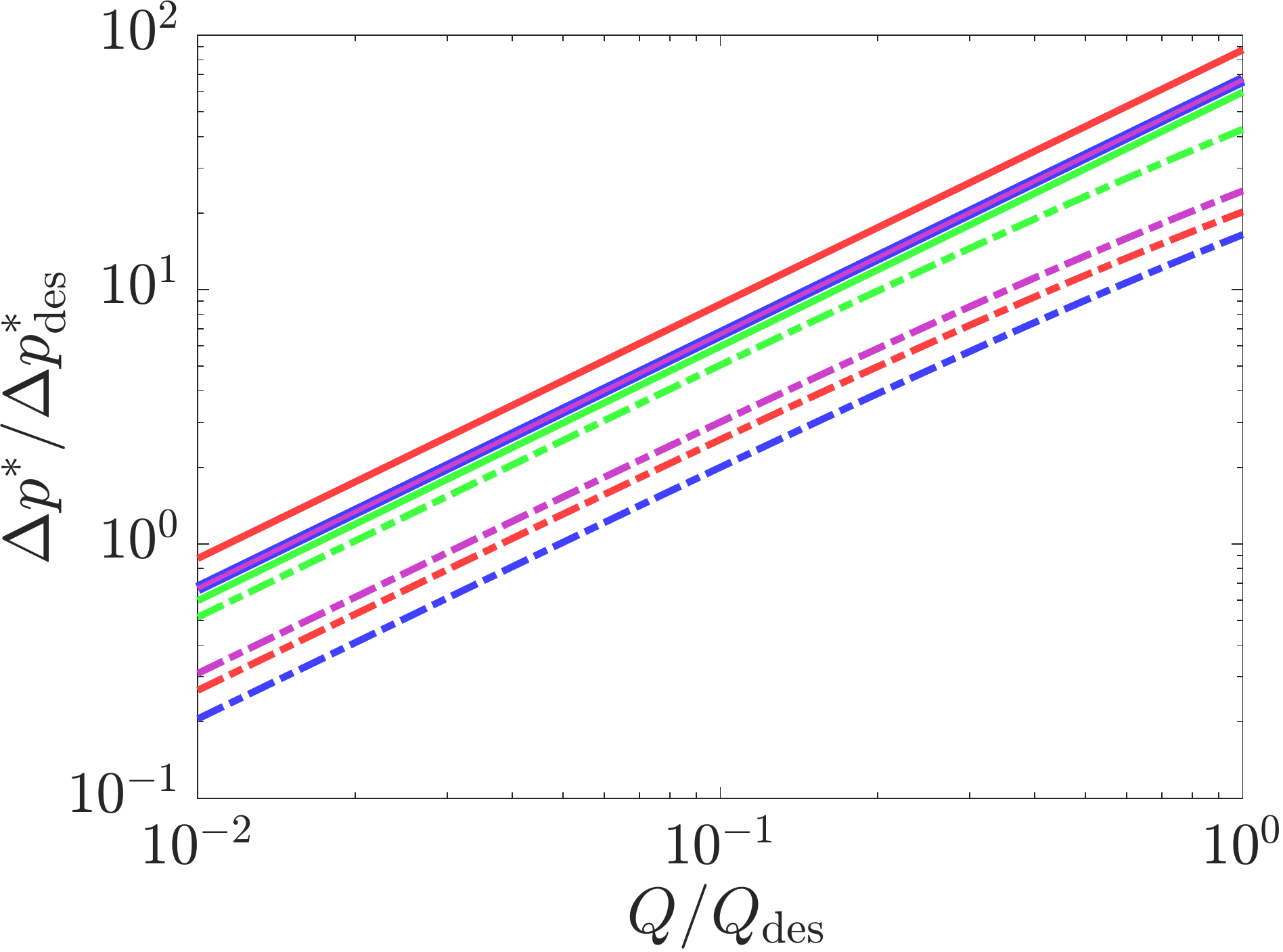}\label{subfig:QdP_Q_slip}}\qquad
\subfloat[]{\includegraphics[width=0.3\textwidth]{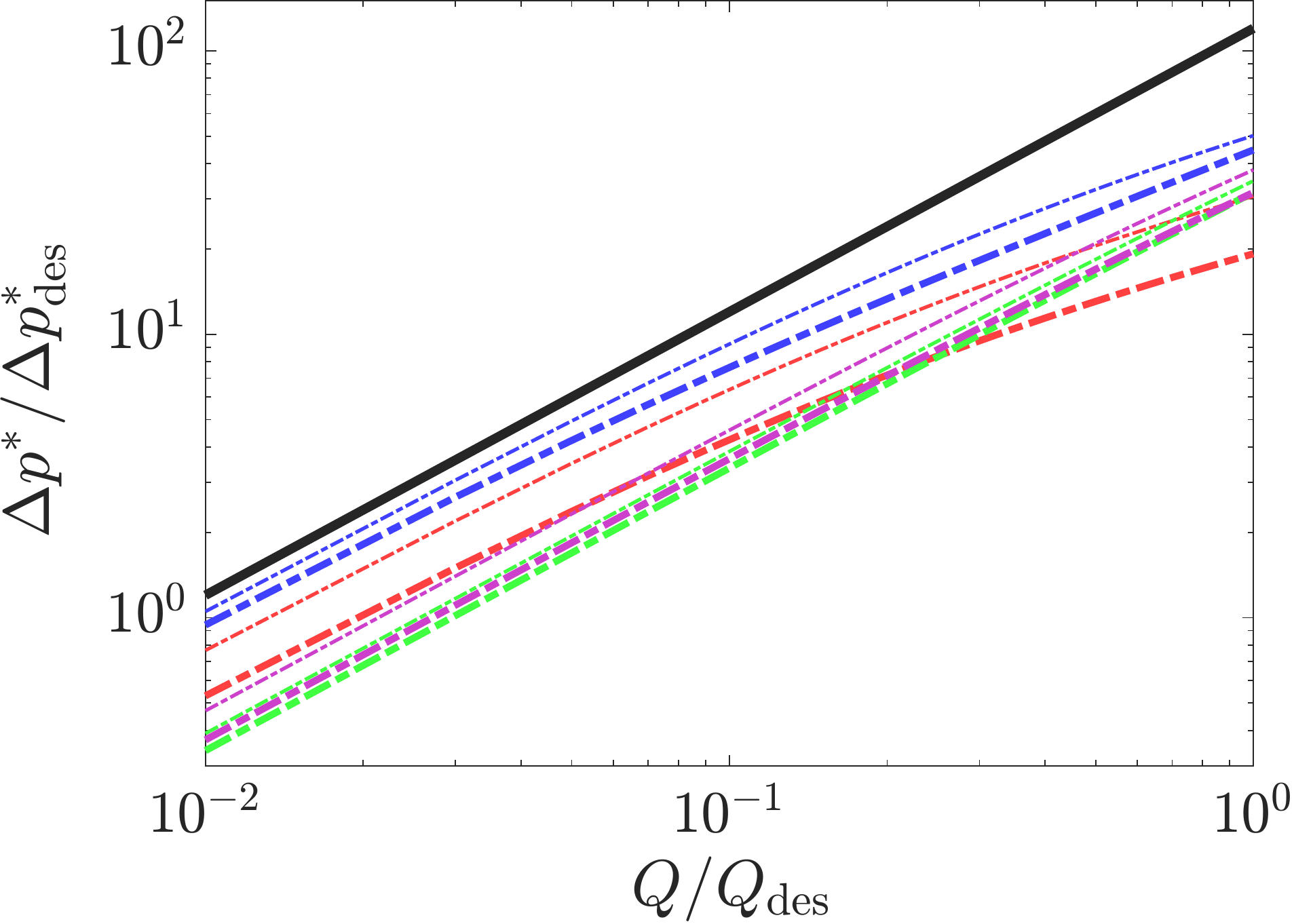}\label{subfig:QdP_Q_taper}}\qquad
\subfloat[]{\includegraphics[width=0.3\textwidth]{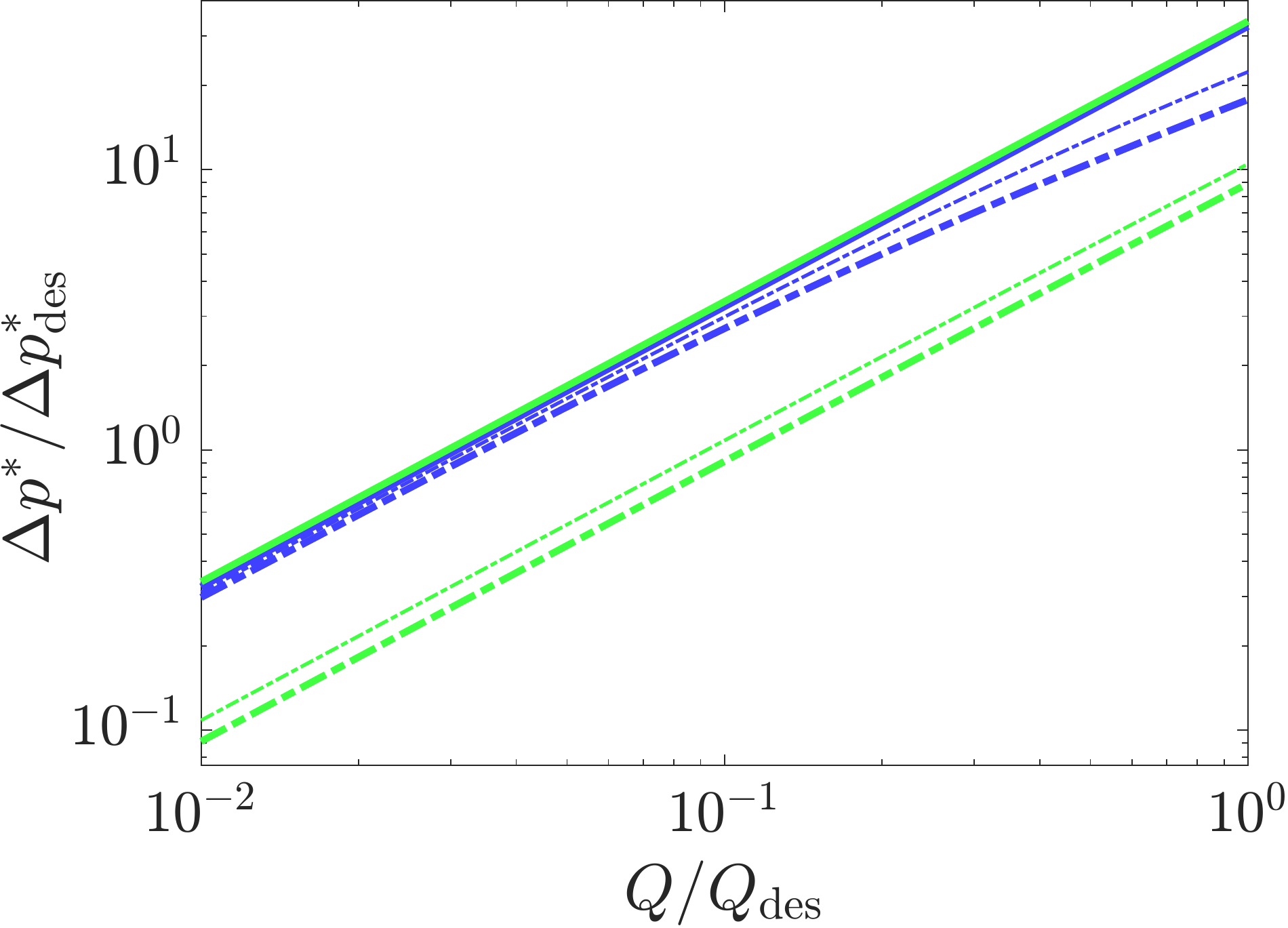}\label{subfig:QdP_Q_slipsh}}
\caption{(Color online.) Normalized pressure drop, $\Delta p^*/\Delta p_{\rm des}^*$ (where $\Delta p_{\rm des}^* = (\kappa/\gamma^3)(\mu Q_{\rm des}/L^2)$), versus normalized flow rate, $Q/Q_{\rm des}$, as the flow rate is increased from a hundredth of the design flow rate ($Q=0.01Q_{\rm des}$) to the design flow rate ($Q=Q_{\rm des}$); we have defined $\Delta p^*$ as $\Delta p^* = p^*|_{x=-{1}/{\kappa}}-p^*|_{x={1}/{\kappa}}$; panels (a,b) correspond to panels (a,b) of figures \ref{fig:slow} to \ref{fig:wave};  panel (c) corresponds to figure \ref{fig:shear}; in (b), all solid curves overlap and are represented by the black curve; the legend above the figure associates the channel shape to color combinations, as well as the solution categories to the linetype combinations used.}
\label{fig:QdP_Q}
\end{figure*}
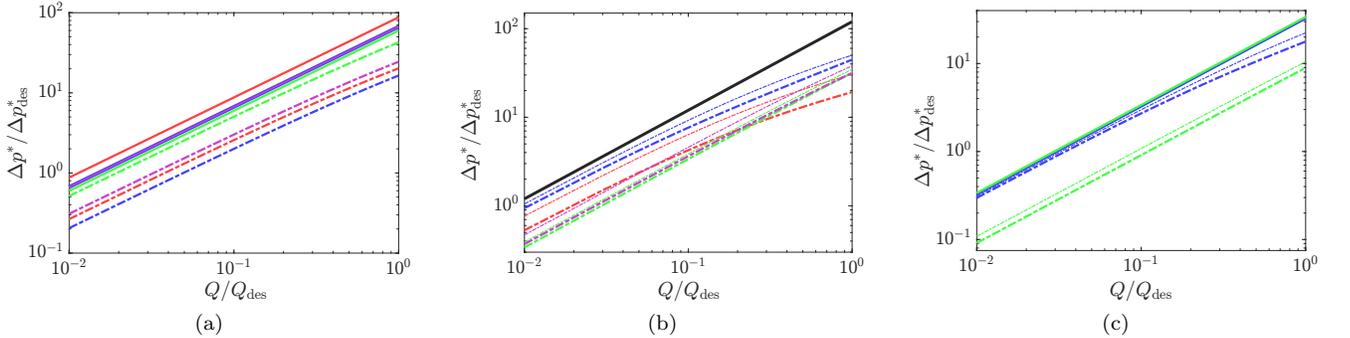

\subsection{Implementation of Obtained Solid Layer Profiling and Slip Length Patterning}\label{subsec:extent}

In figure \ref{fig:sweep_taper}, we present the parametric variation of deformed shape of the fluid domain's top boundary, for the converging-diverging shape (figure \ref{fig:dip}). In figure \subref*{subfig:sweep_slip}, we show the shape change with increasing slip magnitude ($b_0/H$), as obtained from Inverse Problem A. The slip length has the same $x$-dependence as in figure \subref*{subfig:slip_dip} but the magnitude of the slip is varied. The solid layer profiling is also the same as presented in \subref*{subfig:slip_dip}. We see that augmenting slip leads to a converging-diverging shape, compared to a monotonically decreasing one without slip. However, as is to be expected, the variation is weak; that is, we cannot achieve an arbitrary deformation with only slip length patterning of the bottom rigid wall. 

Meanwhile, in figure \subref*{subfig:sweep_taper}, we show the shape change with increasing extent of solid layer profiling (quantified by $m$). We have applied a solid layer profiling:
\begin{equation}
\label{eq:xi_sweep}
\xi(x) = 1+m[\xi_{\text{obtained}}(x)-1],
\end{equation}
where $\xi_{\text{obtained}}(x)$ is the solution to Inverse Problem B for the converging-diverging shape. We vary $m$ from 0.01 to 1, which is equivalent to implementing the $\xi_{\text{obtained}}$ incrementally on the solid layer. Significant deformation can be achieved for $m\approx 1$. This result indicates that the deformed channel shape is stongly sensitive to variations in $\xi$ and, therefore, a sufficiently accurate implementation of the obtained $\xi$ is needed to recover a deformed fluid domain shape that is appreciably close to the desired one. 

\subsection{Pressure Drop Across the Channel}\label{subsec:pdrop}

For many microscale applications, the pressure drop across the microchannel is one of the key measurables, as it can be determined noninvasively \cite{Ozsun2013}. We can evaluate the pressure drop from our mathematical model. In figure \ref{fig:QdP_Q}, we present dimensionless pressure drop versus flow rate curves for the canonical channel shapes from section~\ref{subsec:canonical}. Dimensional counterparts can be obtained by simply scaling the curves shown.


\begin{table*}[htb]
\def\arraystretch{1.25}
\caption{Dimensional parametric values for two representative systems for the microchannel fluid domain shapes presented in section \ref{sec:results} and appendix \ref{sec:comparison}.}
\label{tab:basic_case}
\begin{ruledtabular}
\begin{tabular}{ll@{\hskip 10pt}|lllp{2in}}
\textbf{Material Property}												&	\textbf{Value}								&	\textbf{Geometry \& Flow}				&	\textbf{1\textsuperscript{st} Rep.\ Syst.}		&			\textbf{2\textsuperscript{nd} Rep.\ Syst.}						\\[3pt]
\hline
$\mu$																	&	$1$ mPa$\cdot$s 				 			&	$L$											&	5 cm 				 						&	50 cm										\\[3pt]
$\rho$																	&	$1$ g/cm$^3$			 					&	$H$ 													&	10 $\mu$m				 					&	100 $\mu$m 									\\[3pt]
$\displaystyle E_{\text{Y}} = \frac{3\lambda G+2G^2}{\lambda+G}$		&	$9.5$ kPa										&	$\max_x\Delta(x)$											&	$\sim$ 1 cm								&	$\sim$ 5 cm									\\[3pt]
$\displaystyle \nu = \frac{\lambda}{2(\lambda+G)}$\rule{0pt}{4ex} 						&	0.46										&	$Q_{\rm des}$													&	$6.25\times 10^{-10}$ m$^3$/s							&	$6.25\times 10^{-8}$ m$^3$/s						\\[4pt]
\end{tabular}
\end{ruledtabular}
\end{table*}

\begin{table*}[htb]
\def\arraystretch{1.25}
\caption{Dimensional parametric values for another two representative systems, having nearly-incompressible solid layer material, for the microchannel fluid domain shapes presented in section \ref{sec:results} and appendix \ref{sec:comparison}.}
\label{tab:basic_case_IC}
\begin{ruledtabular}
\begin{tabular}{ll@{\hskip 10pt}|lllp{2in}}
\textbf{Material Property}												&	\textbf{Value}								&	\textbf{Geometry \& Flow}				&	\textbf{1\textsuperscript{st} Rep.\ Syst.}		&			\textbf{2\textsuperscript{nd} Rep.\ Syst.}						\\[3pt]
\hline
$\mu$																	&	$1$ mPa$\cdot$s 				 			&	$L$											&	5 cm 				 						&	50 cm										\\[3pt]
$\rho$																	&	$1$ g/cm$^3$			 					&	$H$ 													&	10 $\mu$m				 					&	100 $\mu$m 									\\[3pt]
$\displaystyle E_{\text{Y}} = \frac{3\lambda G+2G^2}{\lambda+G}$	&	$9.695$ kPa										&	$\max_x\Delta(x)$											&	$\sim$ 1 cm								&	$\sim$ 5 cm									\\[3pt]
$\displaystyle \nu = \frac{\lambda}{2(\lambda+G)}$\rule{0pt}{4ex} 						&	0.49										&	$Q_{\rm des}$													&	$3.7\times 6.25\times 10^{-10}$ m$^3$/s							&	$3.7\times 6.25\times 10^{-8}$ m$^3$/s						\\[4pt]
\end{tabular}
\end{ruledtabular}
\end{table*}

First, in figure \subref*{subfig:QdP_Q_slip}, we present the situations for which slip length patterning leads to the desired channel shapes (panels (a) of figures \ref{fig:slow} to \ref{fig:dip}). It is clear that the pressure drop is significantly smaller for deformable channels (dashed-dotted curves) than for rigid ones (solid curves). This is the expected situation (see, e.g., \cite{Gervais2006}) as the channel can bulge by 40--80\% (see panels (a) of figures \ref{fig:slow} to \ref{fig:dip}). 
Second, in figure \subref*{subfig:QdP_Q_taper}, we present the situations for which solid layer profiling is used to obtain the desired channel shapes (panels (b) of figures \ref{fig:slow} to \ref{fig:dip}). Again, softness of the top wall leads to much smaller pressure drop compared to the rigid case, which is again expected because the channel can bulge by as much as 200\% (see panels (b) of figures \ref{fig:slow} to \ref{fig:dip}). The pressure drop for a rigid top wall, represented by the black solid line, follows the relation:
\begin{equation}
\label{eq:pdrop_dim}
\Delta p^*_{\rm rigid} = \frac{24\mu QL}{H^3} \quad \implies \quad \Delta p_{\rm rigid} = \frac{24}{\kappa}.
\end{equation}
Lastly, in figure \subref*{subfig:QdP_Q_slipsh}, we present the pressure drop--flow rate curves for the situations in which solid layer profiling is used to obtain the desired channel shapes, while slip length patterning is used to control the bottom wall shear rate (figure \ref{fig:shear}). The inferences are identical to those discussed regarding panels (a) and (b).

From the analysis in this subsection, we highlight two key conclusions. First, no clear pattern or correlation emerges between the pressure drops for the four channel shapes. Second, for any of the canonical shapes studied, the trends are distinct for each of the three situations --- rigid solid layer, constant-thickness solid layer, and profiled solid layer. Both these inferences are an outcome of the individual nature of the complex interplay of slip length patterning, solid layer profiling and imposed outlet pressure. Nevertheless, as shown in figure \ref{fig:QdP_Q}, the the pressure drop (a practical figure of merit for microsystem design) can be quantified by the mathematical framework proposed in our study, specifically equation \eqref{eq:ode_p1}, which captures the key details of this coupled multiphysics problem.

\section{Conclusion}\label{sec:conclusion}
In this study, we have presented a theoretical framework for designing passive control of the shape of a the flow conduit inside of a compliant microchannel. Specifically, we demonstrated how the fluid--soft solid interface, between the fluid domain and a compliant wall coating, can be tuned under flow in a `slit' setup commonly used in modeling. To this end, we modeled the top wall of the microchannel as a soft coating of given axially-varying thickness, attached to a rigid platform above it. The variation of thickness, and its coupling to the viscous fluid flow via low Reynolds number fluid--structure interaction \cite{DS14}, allowed us to tune the flow conduit passively. Additionally, we incorporated patterned hydrodynamic slip along the rigid bottom wall, which is commonly used in microfluidics (enabled by nano-patterning of the channel surfaces), to manipulate the magnitude of the shear rate in the flow. The latter is desirable when dealing with cells and biofluids in labs-on-a-chip \cite{Sebastian2018}. Within the broad context of passive shape control, we presented `Inverse Problems,' in which we fed-in the desired axial upper wall shape variation and the bottom wall axial shear rate variation as inputs, and then we solved for a suitable solid layer profile and a slip length pattern. Specifically, these calculations were enabled by the central mathematical result of this work, equation~\eqref{eq:ode_p1}, which connects the hydrodynamic pressure, the fluid--solid interface deformation, and the slip length variation, all in a single ordinary differential equation.

Using this passive elastohydrodynamics (EHD) approach, we demonstrated several categories of possible axial fluid--solid interface shapes that can be achieved under steady flow, starting from a profiled (but undeformed) configuration. The first pair of canonical shapes have decreasing channel height in the flow-wise direction. Specifically, the first shape exhibits faster decrease in channel height (i.e., larger axial gradient of channel height) near the inlet and outlet, and slower decrease in channel height  (i.e., smaller axial gradient of channel height) in the middle. The second canonical shape exhibits slower decrease in channel height near the inlet and outlet, and faster decrease in channel height in the middle. In particular, the second shape is useful for applications in which flow-focussing is desired \cite{Leshansky2007}. The third type of canonical shape presented is converging-diverging, which can find applications in rheological studies \cite{Zografos2016}. The fourth type of canonical shape exhibits a diverging shape near the inlet and outlet and a converging shape in the middle. Using slip length patterning at the bottom wall, we controlled the shear rate magnitude at the bottom wall for the `flow-focusing' and the `converging-diverging' shapes. 

Since our model is dimensionless, it follows that the canonical shapes described above can arise in a multitude of practical systems, in terms of system geometry, flow rate and material properties. As an illustration, in each of tables \ref{tab:basic_case} and \ref{tab:basic_case_IC}, we present two representative physical systems, separated in geometric scale by one order of magnitude, for which the solutions presented in this study would be applicable.

Having showcased passive EHD (termed approach (iii) in section \ref{sec:intro} and the focus of this study) as a useful design tool for controlling microchannel conduit shape under flow, we highlight two disadvantages it has in comparison to active EHD (termed approach (ii) in section \ref{sec:intro}). First, implementation of passive EHD requires an \textit{a priori} theoretical analysis because the approach is `hands-off' during operation (it should however be noted that implementation of active EHD can also require significant efforts in calibrating the force--deformation relationships \cite{Boyko2018}). Second, on a per-setup basis, passive EHD offers less versatility compared to active EHD --- while active EHD allows actuating a particular microchannel setup to obtain multiple kinds of shapes, passive EHD enables only a pre-specified shape designing \textit{a priori}. In this study, we have provide the theory for passive EHD, thus it is now available as an alternative to active EHD in situation in which the latter is not available. Indeed, it might also be of interest to couple the two approaches in the future.

Our passive EHD theory can be used to obtain pre-determined fluid domain shapes in compliant microchannels. These shapes can find applications in multiple areas of microfludics research. An example is the use of converging-diverging channel shapes towards replicating extensional flows \cite{Zografos2016} (as discussed at the end of section \ref{subsec:canonical} as well) for studies on, e.g., dynamics of DNA during flow focusing \cite{Wong2003}. This shape can also be utilized to obtain a desired extent of constriction in the flow conduit, finding use for mimicking stenosis \cite{Venugopal2018} and manipulating polydispersity of droplet generation using softness of the microchannel \cite{Pang2014}. On the other hand, the diverging-converging-diverging geometry can be attuned further to obtain slow divergence followed by sudden convergence and subsequently sudden divergence again \cite{Yaginuma2013} or even a chain of such configurations  \cite{Rodrigues2016}, which can find use in studies of red blood cell dynamics in blood vessels. The converging (fast and slow) channel shapes can be utilized to optimize cell capture and release \cite{Gurkan2011} and particle trapping strategies \cite{Chen2019}. Extensions of this work could include consideration of axial variation of the soft coating's elasticity parameters \cite{Karan2020b}, thick structures and the absence of a platform on top of the solid layer \cite{Wang2019}, shear-dependent viscosity of a non-Newtonian fluid passing through the conduit \cite{Boyko2017,Anand2019}, electrokinetic effects and non-hydrodynamic forces \cite{Karan2019}, amongst others.

\section*{Acknowledgements}
I.C.C.\ thanks the Department of Mechanical Engineering at IIT Kharagpur for its hospitality during his visit there in December 2019, during which this work was initiated. The visit and this research was enabled by the Scheme for Promotion of Academic and Research Collaboration (SPARC), a Government of India Initiative, under Project Code SPARC/2018-2019/P947/SL. I.C.C.\ additionally received partial support from the US National Science Foundation under grant No.\ CBET-1705637. We thank Xiaojia Wang for a careful reading of the manuscript.


\begin{appendix}

\section{Governing Equations}\label{subsec:gdesbcs}

The flow in the fluid domain satisfies the continuity equation,
\begin{equation}
\label{eq:continuity}
\frac{\partial v_x^*}{\partial x^*}+\frac{\partial v_y^*}{\partial y^*}=0,
\end{equation}
and the steady 2D incompressible Navier--Stokes equations, 
\begin{equation}
\label{eq:xmom}
\rho\left(v_x^*\frac{\partial v_x^*}{\partial x^*}+v_y^*\frac{\partial v_x^*}{\partial y^*}\right)=-\frac{\partial p^*}{\partial x^*}+\mu\left(\frac{\partial^2 v_x^*}{\partial x^{*2}}+\frac{\partial^2 v_x^*}{\partial y^{*2}}\right),
\end{equation}
\begin{equation}
\label{eq:ymom}
\rho\left(v_x^*\frac{\partial v_y^*}{\partial x^*}+v_y^*\frac{\partial v_y^*}{\partial y^*}\right)=-\frac{\partial p^*}{\partial y^*}+\mu\left(\frac{\partial^2 v_y^*}{\partial x^{*2}}+\frac{\partial^2 v_y^*}{\partial y^{*2}}\right).
\end{equation}
The system is closed by the Navier slip and no-penetration boundary condition at the rigid bottom wall:
\begin{align}
\label{eq:NavierSlipbottom}
\left.v_x^*\right|_{y^*=0}&=\left.b(x^*)\frac{\partial v_x^*}{\partial y^*}\right|_{y^*=0},\\
\label{eq:nopenbottom}
\left.v_y^*\right|_{y^*=0}&=0,
\end{align}
respectively, in addition to the no-slip and no-penetration boundary condition at the deformed fluid--solid interface:
\begin{equation}
\label{eq:noslipnopentop}
\left.v_x^*\right|_{y^*=H-h^*(x^*)}=\left.v_y^*\right|_{y^*=H-h^*(x^*)}=0.
\end{equation}
The pressure at the outlet is imposed:
\begin{equation}
\label{eq:outletp}
\left.p^*\right|_{x^* = L}=p_0^*,
\end{equation}
and the flow rate of $Q$ per unit width, at any given cross-section of the channel, is
\begin{equation}
\label{eq:flowconst}
\int_{y^*=0}^{y^*=H-h^*(x^*)}v_x^*\,dy^* = Q.
\end{equation}

The deformation of the soft elastic layer (the solid domain) is governed by the elastostatic equilibrium equations:
\begin{equation}
\label{eq:mecheqcompact}
\nabla^*\cdot\underline{\underline{\sigma}}^* = \vec{0}.
\end{equation}
The solid's Cauchy stress tensor is given in terms of the displacement gradient as,
\begin{equation}
\label{eq:substratestress}
\underline{\underline{\sigma}}^* = \lambda\left(\nabla^*\cdot\vec{u}^*\right)\underline{\underline{I}}+G\left(\nabla^*\vec{u}^*+(\nabla^*\vec{u}^*)^\top\right),
\end{equation}
where $\underline{\underline{I}}$ is the identity tensor and a $\top$ superscript denotes the transpose. 
Substituting equation \eqref{eq:substratestress} into equation \eqref{eq:mecheqcompact} yields the two components of the mechanical equilibrium equation:
\begin{subequations}\begin{align}
\label{eq:mecheq_x}
(\lambda+2G)\frac{\partial^2 u_x^*}{\partial x^{*2}}+G\frac{\partial^2 u_x^*}{\partial \bar{y}^{*2}} + (\lambda+G)\frac{\partial^2 u_{\bar{y}}^*}{\partial x^* \partial \bar{y}^*} &= 0,\\
\label{eq:mecheq_y}
G\frac{\partial^2 u_{\bar{y}}^*}{\partial x^{*2}}+(\lambda+2G)\frac{\partial^2 u_{\bar{y}}^*}{\partial \bar{y}^{*2}} + (\lambda+G)\frac{\partial^2 u_x^*}{\partial x^* \partial \bar{y}^*} &= 0.
\end{align}\label{eq:Cauchy}\end{subequations}
This system is closed by zero-displacement boundary conditions at the solid-platform interface:
\begin{equation}
\label{eq:nodisptop}
\left.u_x^*\right|_{\bar{y}^* = \Delta(x^*)}=\left.u_{\bar{y}}^*\right|_{\bar{y}^* = \Delta(x^*)}=0,
\end{equation}
and the traction balance condition at the fluid--solid interface,
\begin{equation}
\label{eq:tracbalcompact}
\underline{\underline{\sigma}}^*\cdot\hat{n}^* = \underline{\underline{\sigma}}^{f*}\cdot\hat{n}^*.
\end{equation}

Here, $\underline{\underline{\sigma}}^{f*}$ is the Newtonian fluid's stress tensor:
\begin{equation}
\label{eq:fluidstress}
\underline{\underline{\sigma}}^{f*} = -p^*\underline{\underline{I}}+\mu\left(\nabla^*\vec{v}^*+(\nabla^*\vec{v}^*)^\top\right),
\end{equation}
and $\hat{n}^*$ is the normal to the fluid--solid interface:
\begin{equation}
\label{eq:normal}
\hat{n}^* = \frac{\partial h^*}{\partial x^*}\hat{i}+\hat{j}.
\end{equation}
Thus, the traction balance condition, at $y^* = H-h^*(x^*)$ and $\bar{y}^* = -h^*(x^*)$, can expressed in component form as
\begin{subequations}\begin{multline}
\label{eq:tracbal_x}
\left((\lambda+2G)\frac{\partial u_x^*}{\partial x^*}+\lambda\frac{\partial u_{\bar{y}}^*}{\partial \bar{y}^*}\right)\frac{\partial h^*}{\partial x^*}  +G\left(\frac{\partial u_x^*}{\partial \bar{y}^*}+\frac{\partial u_{\bar{y}}^*}{\partial x^*}\right)  \\  = \left(-p^*+2\mu\frac{\partial v_x^*}{\partial x^*}\right)\frac{\partial h^*}{\partial x^*}+\mu\left(\frac{\partial v_x^*}{\partial y^*}+\frac{\partial v_y^*}{\partial x^*}\right),
\end{multline}
\begin{multline}
\label{eq:tracbal_y}
G\left(\frac{\partial u_x^*}{\partial \bar{y}^*}+\frac{\partial u_{\bar{y}}^*}{\partial x^*}\right)\frac{\partial h^*}{\partial x^*}+(\lambda+2G)\frac{\partial u_{\bar{y}}^*}{\partial \bar{y}^*}+\lambda\frac{\partial u_x^*}{\partial x^*} \\ = -p^*+2\mu\frac{\partial v_y^*}{\partial y^*}+\mu\left(\frac{\partial v_x^*}{\partial y^*}+\frac{\partial v_y^*}{\partial x^*}\right)\frac{\partial h^*}{\partial x^*}.
\end{multline}\label{eq:tracbal}
\end{subequations}

\section{Dimensionless Governing Equations}\label{subsec:gdesbcs_nd}

\paragraph{Fluid domain:}
\begin{equation}
\label{eq:continuity_nd1}
\frac{\partial v_x}{\partial x}+\frac{\partial v_y}{\partial y}=0,
\end{equation}

\begin{multline}
\label{eq:xmom_nd1}
\frac{\gamma}{\kappa}\frac{\rho Q}{\mu}\left(v_x\frac{\partial v_x}{\partial x}+v_y\frac{\partial v_x}{\partial y}\right)\\
=-\frac{\partial p}{\partial x}+\frac{\partial^2 v_x}{\partial y^{2}}+\frac{\gamma^2}{\kappa^2} \frac{\partial^2 v_x}{\partial x^{2}},
\end{multline}
\begin{multline}
\label{eq:ymom_nd1}
\frac{\gamma^3}{\kappa^3}\frac{\rho Q}{\mu}\left(v_x\frac{\partial v_y}{\partial x}+v_y\frac{\partial v_y}{\partial y}\right)\\
=-\frac{\partial p}{\partial y}+\frac{\gamma^2}{\kappa^2}\left(\frac{\partial^2 v_y}{\partial y^{2}}+\frac{\gamma^2}{\kappa^2}\frac{\partial^2 v_y}{\partial x^{2}}\right),
\end{multline}
subject to:
\begin{align}
\label{eq:NavierSlipbottom_nd1}
\left.v_x\right|_{y=0}&=\left.\frac{b(x)}{\gamma L}\frac{\partial v_x}{\partial y}\right|_{y=0},\\
\label{eq:nopenbottom_nd1}
\left.v_y\right|_{y=0}&=0,
\end{align}
and
\begin{equation}
\label{eq:noslipnopentop_nd1}
\left.v_x\right|_{y=1-{\phi_0 h(x)}/{\gamma}}=\left.v_y\right|_{y=1-{\phi_0 h(x)}/{\gamma}}=0,
\end{equation}
as well as,
\begin{align}
\label{eq:outletp_nd}
\left.p\right|_{x = 1/\kappa}&=\frac{\gamma^3}{\kappa}\frac{p_0^* L^2}{\mu Q} = \bar{p}_0,\\
\label{eq:flowconst_nd}
\int_{y=0}^{y=1-{\phi_0 h(x)}/{\gamma}}v_x\,dy &= 1.
\end{align}

\paragraph{Solid Domain:}
\begin{align}
\label{eq:mecheq_x_nd1}
\frac{\partial^2 u_x}{\partial \bar{y}^2}+\frac{\beta}{\gamma}\left(1+\frac{\lambda}{G}\right)\frac{\partial^2 u_{\bar{y}}}{\partial x \partial \bar{y}}+\frac{\beta^2}{\gamma^2}\left(2+\frac{\lambda}{G}\right)\frac{\partial^2 u_x}{\partial x^2} &= 0,\\[1mm]
\label{eq:mecheq_y_nd1}
\left(2+\frac{\lambda}{G}\right)\frac{\partial^2 u_{\bar{y}}}{\partial \bar{y}^2}+\frac{\beta}{\gamma}\left(1+\frac{\lambda}{G}\right)\frac{\partial^2 u_x}{\partial x \partial \bar{y}}+\frac{\beta^2}{\gamma^2}\frac{\partial^2 u_{\bar{y}}}{\partial x^2} &= 0,
\end{align}
subject to 
\begin{equation}
\label{eq:nodisptop_nd1}
\left.u_x\right|_{\bar{y} = \xi(x) = {\Delta(x)}/{\beta L}}=\left.u_{\bar{y}}\right|_{\bar{y} = \xi(x) = {\Delta(x)}/{\beta L}}=0,
\end{equation}
as well as
\begin{widetext}
\begin{multline}
\label{eq:tracbal_x_nd1}
\frac{\partial u_x}{\partial \bar{y}}+\frac{\beta}{\kappa}\left[\frac{\partial u_{\bar{y}}}{\partial x}+\left(2+\frac{\lambda}{G}\right)\frac{\phi_0}{\kappa}\frac{\partial u_x}{\partial x}\frac{\partial h}{\partial x}+\frac{\lambda}{G}\frac{\phi_0}{\beta}\frac{\partial u_{\bar{y}}}{\partial \bar{y}}\frac{\partial h}{\partial x} \right] =  -\frac{\beta\kappa}{\gamma^3\phi_0}\frac{\mu Q}{(\lambda+2G)L^2}\left(2+\frac{\lambda}{G}\right) \\ \times \left[-\frac{\phi_0}{\kappa}p\frac{\partial h}{\partial x}+\frac{\gamma}{\kappa}\left(\frac{\partial v_x}{\partial y}+\frac{\gamma^2}{\kappa^2}\frac{\partial v_y}{\partial x}\right)+\frac{\phi_0}{\kappa}\frac{\gamma^2}{\kappa^2}\frac{\partial v_x}{\partial x}\right], \\
\text{at }y = 1-\frac{\phi_0 h(x)}{\gamma},\quad \bar{y} = -\frac{\phi_0 h(x)}{\beta}\approx 0,
\end{multline}
and
\begin{multline}
\label{eq:tracbal_y_nd1}
\frac{\partial u_{\bar{y}}}{\partial \bar{y}}+\frac{\beta}{\kappa}\left[\left(\frac{\lambda}{\lambda+2G}\right)\frac{\partial u_x}{\partial x} + \left(\frac{G}{\lambda+2G}\right)\left(\frac{\phi_0}{\kappa}\frac{\partial u_{\bar{y}}}{\partial x}+\frac{\phi_0}{\beta}\frac{\partial u_x}{\partial \bar{y}}\right)\frac{\partial h}{\partial x}\right]  \\
= - \frac{\beta\kappa}{\gamma^3\phi_0}\frac{\mu Q}{(\lambda+2G)L^2} \left[p-\frac{\gamma}{\kappa}\left\{\frac{2\gamma}{\kappa}\frac{\partial v_y}{\partial y}+\frac{\phi_0}{\kappa}\left(\frac{\partial v_x}{\partial y}+\frac{\gamma^2}{\kappa^2}\frac{\partial v_y}{\partial x}\right)\right\}\right], \\
\text{at }y = 1-\frac{\phi_0 h(x)}{\gamma},\quad \bar{y} = -\frac{\phi_0 h(x)}{\beta}\approx 0.
\end{multline}


\section{Newton--Raphson Method for Equation~\eqref{eq:ode_p1}}\label{sec:Newton}

We discretize the $x$-axis into $n$ points as $x_i = -1/\kappa + 2(i-1)/\big((n-1)\kappa\big)$, $i=1,\hdots,n$. We used $n$ = 1001 points in the implementation of the numerical scheme, having verified this number is sufficient to ensure grid-independence of solutions. The components $R_i$ of the residual vector $\vec{R}$ and $J_{i,j}$ of the requisite Jacobian matrix $\underline{\underline{J}}$ are defined as:
\begin{equation}
\label{eq:Res}
\begin{split}
R_i &=  \left(1+\frac{\phi_0 p_i}{\bar{\phi}_i\gamma}\right)^3\left(1+\frac{\phi_0 p_i}{\bar{\phi}_i\gamma}+\frac{4b_i}{\gamma L}\right)^3\sum_{j=0}^{j=2} a_{i,j}^{(\text{FD})}p_{i+j} + 12\left(1+\frac{\phi_0 p_i}{\bar{\phi}_i\gamma}+\frac{b_i}{\gamma L}\right),\quad i=1;\\
R_i &=  \left(1+\frac{\phi_0 p_i}{\bar{\phi}_i\gamma}\right)^3\left(1+\frac{\phi_0 p_i}{\bar{\phi}_i\gamma}+\frac{4b_i}{\gamma L}\right)^3\sum_{j=0}^{j=2} a_{i,j}^{(\text{CD})}p_{i+j} +12\left(1+\frac{\phi_0 p_i}{\bar{\phi}_i\gamma}+\frac{b_i}{\gamma L}\right),\quad i=2~\text{to}~n-1;\\
R_i &=  p_i-\frac{\gamma^3}{\kappa}\frac{p_0^* L^2}{\mu Q},\quad i=n;
\end{split}
\end{equation}
\begin{equation}
\label{eq:Jac}
\begin{split}
J_{i,i+j} &= \left(1+\frac{\phi_0 p_i}{\bar{\phi}_i\gamma}\right)^3\left\{\frac{4\phi_0}{\bar{\phi_i}\gamma}\left[\left(\frac{1+\frac{\phi_0 p_i}{\bar{\phi}_i\gamma}+\frac{3b_i}{\gamma L}}{1+\frac{\phi_0 p_i}{\bar{\phi}_i\gamma}}\right)  \sum_{j=0}^{j=2} a_{i,j}^{x(\text{CD})}p_{i+j}+3\right]\delta_{i,j}+ \left(1+\frac{\phi_0 p_i}{\bar{\phi}_i\gamma}+\frac{4b_i}{\gamma L}\right)a_{i,j}^{(\text{CD})}\right\};\\
&\phantom{=}\hspace{350pt} i=1,\\
J_{i,i+j} &= \left(1+\frac{\phi_0 p_i}{\bar{\phi}_i\gamma}\right)^3\left\{\frac{4\phi_0}{\bar{\phi_i}\gamma}\left[\left(\frac{1+\frac{\phi_0 p_i}{\bar{\phi}_i\gamma}+\frac{3b_i}{\gamma L}}{1+\frac{\phi_0 p_i}{\bar{\phi}_i\gamma}}\right)\sum_{j=0}^{j=2} a_{i,j}^{x(\text{FD})}p_{i+j}+3\right]\delta_{i,j}+\left(1+\frac{\phi_0 p_i}{\bar{\phi}_i\gamma}+\frac{4b_i}{\gamma L}\right)a_{i,j}^{(\text{FD})}\right\};\\
& \phantom{=}\hspace{300pt}  i=1,~j=0~\text{to}~2,
\\
J_{i,i} &=1,\qquad i=n.
\end{split}
\end{equation}
\end{widetext}
where $p_i \approx p(x_i)$, $\bar{\phi}_i \approx \bar{\phi}(x_i)$ and $b_i\approx b(x_i)$ are the nodal value approximations. Here, $a^{\text{s}}_{i,j}$ represents the coefficients for the node $i+j$ corresponding to the finite-difference approximation of the derivative $d/dx$ at the node $i$, the superscript `s' is `CD' for central-difference and `FD' for forward difference; $\delta_{i,j}$ is the Kronecker-delta symbol, i.e., $\delta_{i,j}= 1$ when $j=i$ and $\delta_{i,j}=0$ when $j\ne i$. 

To obtain the numerical solution, we iterate 
$\vec{p} \mapsto \vec{p}-\underline{\underline{J}}^{-1}\vec{R}$ until $\|\vec{R}\|$ becomes smaller than a prescribed tolerance, which we took to be $10^{-5}$.


\section{Exploring the Validity of the 2D Model}\label{sec:comparison}

\begin{figure*}[!htb]
\centering
\subfloat[]{\includegraphics[width=0.4\textwidth]{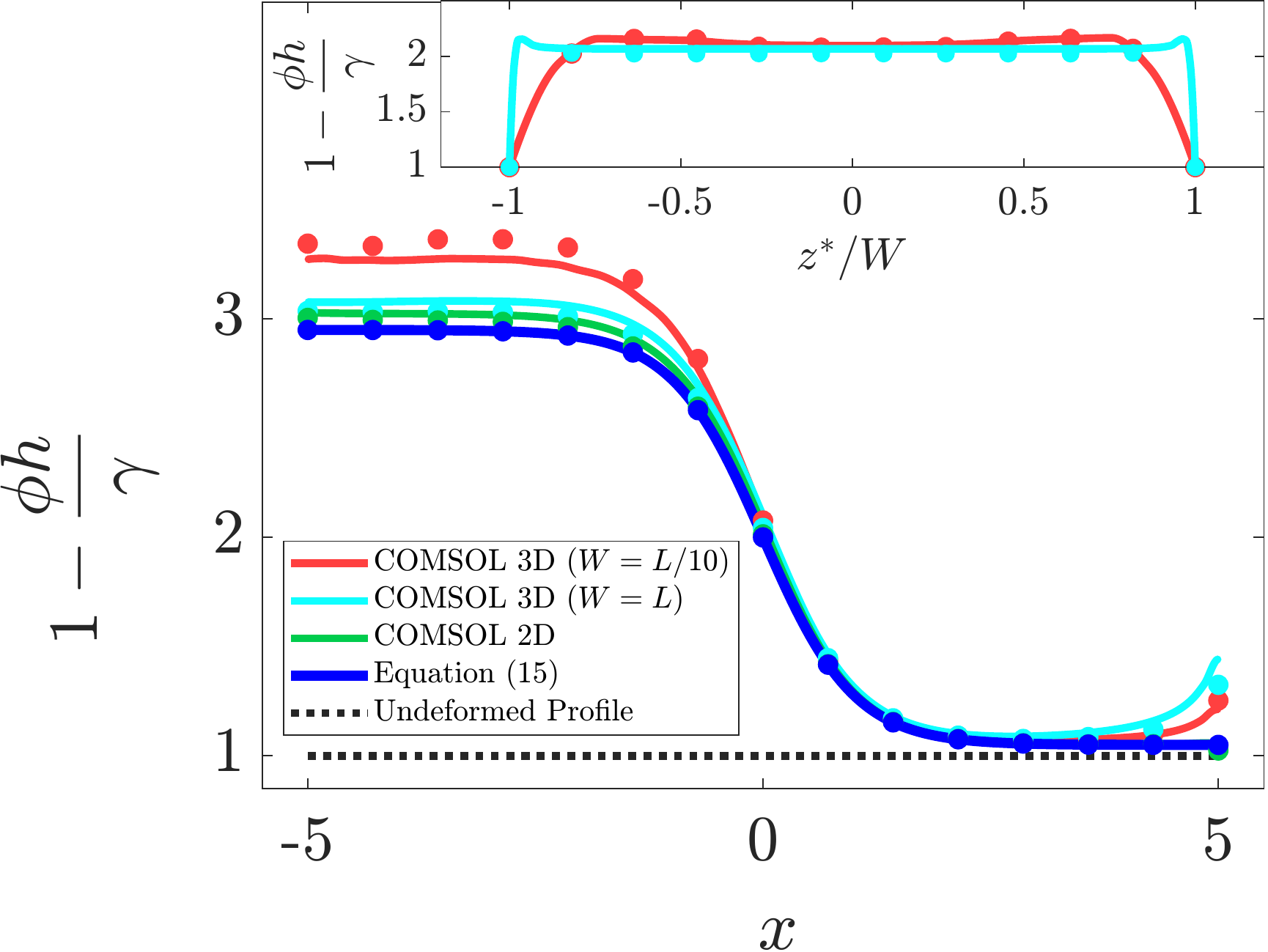}\label{subfig:fast_gap_compare}}\hspace{22.5pt}
\subfloat[]{\includegraphics[width=0.4\textwidth]{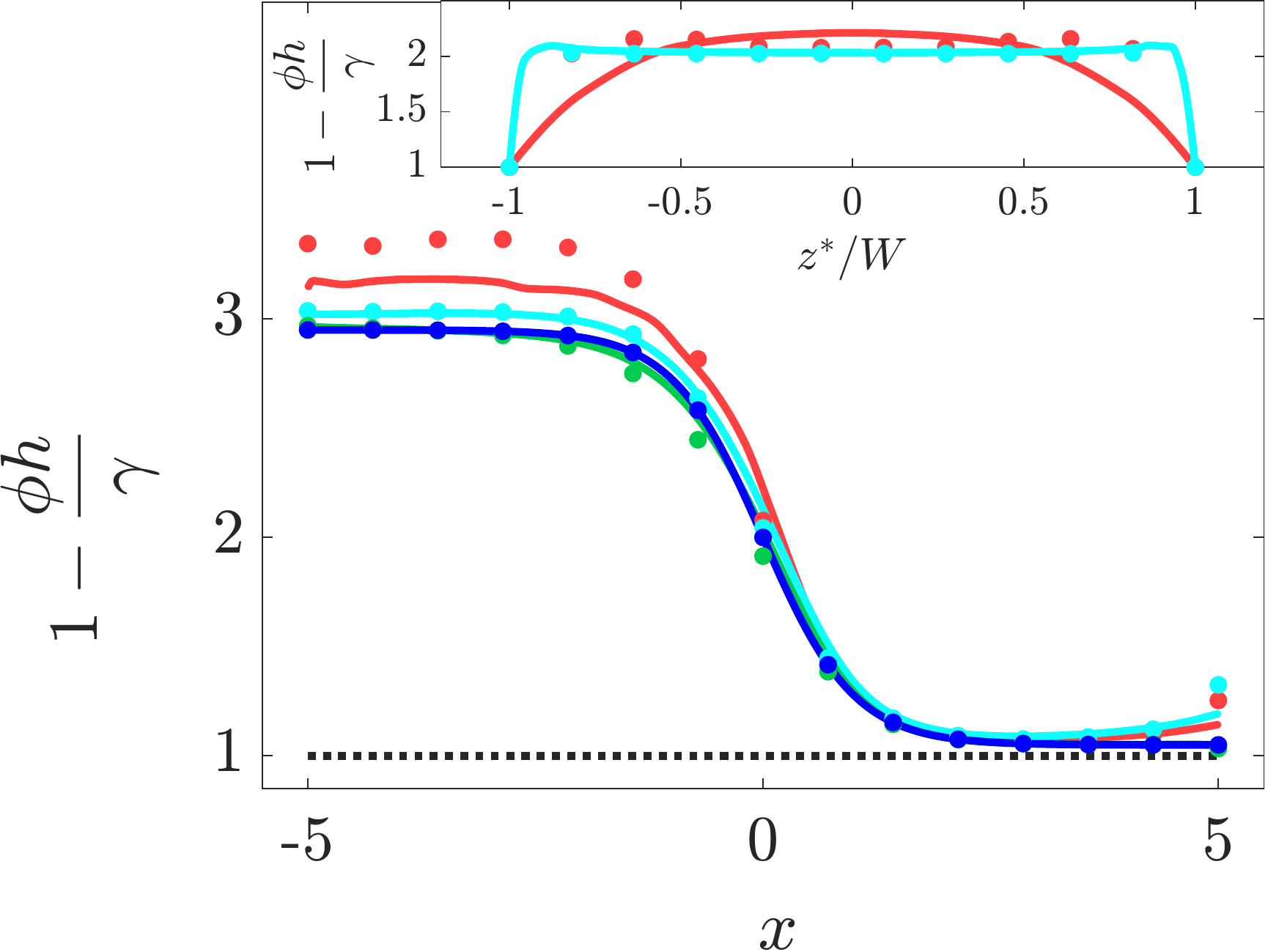}\label{subfig:fast_sh_gap_compare}}
\caption{(Color online.) Dimensionless deformed shape of the fluid domain as obtained from 3D COMSOL simulations, 2D COMSOL simulations, and the solution of equation \eqref{eq:ode_p1}. Panels (a) and (b) correspond to figures \protect\subref*{subfig:taper_fast} and \protect\subref*{subfig:taper_fast_sh}, respectively. The legend in (a) applies to both panels. In each panel, the solid curves and circle markers represent the solution based on parameter values in tables \ref{tab:basic_case} and \ref{tab:basic_case_IC}, respectively. For the 3D COMSOL simulations, the axial variation of the gap height (presented as the main plot in the panels) is at the midplane $z^*=0$. The insets represent the span-wise dimensionless deformed shape of the fluid domain, across the $x = 0$ plane, from the 3D COMSOL simulations.}
\label{fig:fast_gap_compare}
\end{figure*}

In this study, several simplifications have been made towards reducing the complete system behavior to the single ODE, equation \eqref{eq:ode_p1}. While some of the assumptions are standard to the research area of flows through microchannels (including deformable ones), some assumptions warrant extra examination to ascertain the validity of our model. We identify 3 concerns:
\begin{enumerate}
\item We have considered the setup to be sufficiently wide in the plane perpendicular to the flow, i.e., we have considered a 2D setup. While this `slit' setup is commonly assumed when modeling microchannels \cite{Gervais2006}, the mathematical expression of this assumption, $W \gg L$, where $2W$ is the width `into-the-paper', might not always be realizable.\label{concern1}
\item Aggressive solid layer profiling leads to variations in the solid layer thickness over short axial distances. The characteristic length over which the solid domain varies is then liable to be smaller than its physical axial length. This indicates that the pressure at a point on the fluid--solid interface can affect deflection of not only that point but also its neighboring points. This effect is not captured by the Winkler-like relation (equation \eqref{eq:h}).\label{concern2}
\item We have obtained the Winkler-like relation (equation \eqref{eq:h}) by an asymptotic reduction of the governing equations and boundary conditions for the solid domain. This calculation rests on the scaling assumption that both $u_y$ and $u_x$ scale as $\phi_0 L$, which is admissible only when $\lambda$ and $G$ are of the same scale, i.e., the solid layer is appreciably compressible (see appendix B in \cite{Karan2019} and the fourth paragraph of section II in \cite{Rallabandi2017}). \label{concern3}
\end{enumerate}

Towards addressing these potential limitations, we have conducted 2D as well as 3D COMSOL simulations of the fluid-structure interactions for two situations (which are the most extreme ones in the context of the concerns listed above), corresponding to figures \subref*{subfig:taper_fast} and \subref*{subfig:taper_fast_sh}, which were discussed in section \ref{sec:results}. The consolidated results of these simulations are presented in figure \ref{fig:fast_gap_compare}.

We first address concern \ref{concern1}. To investigate whether $W \gg L$ must be strictly satisfied for the validity of our theory, we have conducted 3D numerical simulations corresponding to the 2D systems from figures \subref*{subfig:taper_fast} and \subref*{subfig:taper_fast_sh}, taking the two situations of $W=L$ and $W=L/10$. We emphasize that these simulations solve the complete 3D elastostatic equations, two-way coupled to the steady incompressible Navier--Stokes equations, without any assumptions of slit-like geometry and slenderness. We have additionally conducted simulations for the 2D system as well (i.e., the slit-like geometry is assumed but slenderness is not assumed). Figure \ref{fig:fast_gap_compare} shows that the numerical solutions (2D simulations and 3D simulations with $W=L$) and the solution of our proposed model all match well. Even the results from the 3D simulations with $W=L/10$ are appreciably close. Furthermore, examining the insets, we observe that, for the 3D systems, the edge effects are restricted to a small region near the edge planes ($z^*=-W$ and $z^*=W$), and the solid layer deformation close to the central plane $z^*=0$ is not appreciably affected by edge effects. Hence, we deduce that our framework captures the solid layer deformation behavior  well for 3D setups, even the ones for which the width is comparable to the axial length. 

Next, concerns \ref{concern2} and \ref{concern3} are about the applicability of the Winkler-like relation, equation \eqref{eq:ode_p1}. The Winkler mattress model is a simple yet effective model, which was designed to study the deformation behavior of beams and plates resting on elastic foundations. In this model, the elastic foundation was \textit{a forteriori} likened to a collection of springs, with each spring responding to the load applied only on its point of attachment with the beam/plate. The spring constant is expected to be a function of the solid layer elastic properties. This model, with some refinements and generalizations, has been found to be fruitful for analyzing many mechanics problems \cite{Dillard2018}. On the other hand, the Winkler-like relation we obtained, although having the same functional form, is derived starting from the complete elastostatic equations, and hence, it is not based on an assumption like the original Winkler mattress model. Nevertheless, the derived Winkler-like relation's validity rests on two conditions --- (a) the solid layer should be slender, and (b) the solid layer should be sufficiently compressible. Thus, concerns \ref{concern2} and \ref{concern3} are really about whether and when conditions (a) and (b), respectively, break down.

Addressing first concern \ref{concern2}, we observe that the slenderness of the solid layer, for our setup, is represented by $\beta/\kappa$. For the cases we have studied in section \ref{sec:results}, the maximum value of $\beta/\kappa$ occurs for the case of figure \subref*{subfig:taper_dip_sh} ($\beta/\kappa \approx 1$). However, as we can see in figure \subref*{subfig:fast_sh_gap_compare}, even for this case, the solution from our framework (blue) and the solution from COMSOL (red and green) match. In other words, our framework is able to represent the solution of the complete elastostatic equations (as found by numerical simulations using COMSOL). Hence, even for $\beta/\kappa \approx 1$, the assumption of slenderness of the solid layer remains admissible.

Addressing concern \ref{concern3} next, we recall that some of the materials commonly used in microfluidics, an example being polydimethylsiloxane (PDMS) \cite{Xia1998,SSA04}, are often considered incompressible, with $\nu \approx 0.499$ being one commonly used value. However, it has long been conjectured that such materials are bound to have some compressibility (``[a] definitive value for the Poisson's ratio ... is not readily available in the literature'' \cite[p.~6]{Johnston2014}), and more recent measurements have suggested that the Poisson's ratio can be as low as $\nu \approx 0.46$ \cite{Dogru2018,Raj2018}. Keeping this fact in mind, we have considered the solid layer to be `sufficiently' compressible, and we took $\nu=0.46$ in table \ref{tab:basic_case} and $\nu=0.49$ in table \ref{tab:basic_case_IC}. 

Here, we observe that a condition for the applicability of the classical Winkler mattress model, which is also expected to apply to  our Winkler-like relation derived as equation \eqref{eq:h}, is that $(1-2\nu)\gtrsim (\beta/\kappa)^2$. This condition is obtained based on scaling principles applied to the complete elastostatic equations --- see last paragraph of section 2 in \cite{Chandler2020} and appendix B in \cite{Karan2019}. For the cases studied in section \ref{sec:results} and with dimensional parameter values taken from either of tables \ref{tab:basic_case} and \ref{tab:basic_case_IC}, the example presented in figure \subref*{subfig:taper_fast} satisfies this condition, whereas the example presented in figure \subref*{subfig:taper_fast_sh} is the most adverse in terms of fulfillment of this condition. Therefore, we have compared the solution from our model with the solution from COMSOL simulations of these two cases in figure \ref{fig:fast_gap_compare}(a,b), respectively. The match between our model and simulations, especially in panel (b), indicates that even when the condition $(1-2\nu)\gtrsim (\beta/\kappa)^2$ is not strictly satisfied, the prediction of equation \eqref{eq:ode_p1} is reasonably close to the numerical solution of the fluid--structure interaction problem. This indicates that our framework can apply to solid layer materials that are close to incompressible. 

However, we caution that for materials that are extremely close to being incompressible, i.e., $(1-2\nu) \ll (\beta/\kappa)^2$, the deflection is expected to be $\mathcal{O}(\beta^2)$ smaller and proportional to the Laplacian of the pressure (see appendix A in \cite{Chandler2020} and appendix B in \cite{Karan2019}). This case was not considered in this study, and is beyond the scope of the present work.

\end{appendix}

\bibliography{refs_short}
\end{document}